\documentclass[12pt]{article}
\usepackage{cite}
\usepackage{hyperref}
\hypersetup{colorlinks,allcolors=blue}

\usepackage{amssymb,braket,slashed,
enumerate,bbold,empheq,xcolor,cases,
siunitx,booktabs,soul, multirow}

\usepackage{amsmath}
\usepackage[displaymath]{lineno}

\makeatletter
\let\LN@align\align
\let\LN@endalign\endalign
\renewcommand{\align}{\linenomath\LN@align}
\renewcommand{\endalign}{\LN@endalign\endlinenomath}
\let\LN@gather\gather
\let\LN@endgather\endgather
\renewcommand{\gather}{\linenomath\LN@gather}
\renewcommand{\endgather}{\LN@endgather\endlinenomath}
\makeatother

\newcounter{starfoot}
\begin{document}

\thispagestyle{empty}
\begin{titlepage}
\boldmath
\begin{center}
  \Large {\bf Analysis of atomic-clock data to constrain variations of fundamental constants}
    \end{center}
\unboldmath
\vspace{0.2cm}
\begin{center}
{
\renewcommand{\thefootnote}{\fnsymbol{starfoot}}
{\large Nathaniel~Sherrill}\footnote{Joint first authors}\renewcommand{\thefootnote}{\arabic{footnote}}\addtocounter{footnote}{-1}\footnote{E-mail: n.sherrill@sussex.ac.uk}$^a$,
{\large  Adam~O.~Parsons}\renewcommand{\thefootnote}{\fnsymbol{starfoot}}\footnotemark{}\renewcommand{\thefootnote}{\arabic{footnote}}\addtocounter{footnote}{-1}\footnote{E-mail: adam.parsons@npl.co.uk}$^b$,
{\large {Charles~F.~A.~Baynham}$^b$},
{\large {William~Bowden}$^b$},
{\large E.~Anne~Curtis}$^b$,
{\large Richard~Hendricks}$^b$,
{\large Ian~R.~Hill}$^b$,
{\large {Richard~Hobson}$^b$},
{\large Helen~S.~Margolis}$^b$,
{\large Billy~I.~Robertson}$^b$,
{\large Marco~Schioppo}$^b$,
{\large Krzysztof~Szymaniec}$^b$,
{\large Alexandra~Tofful}$^b$,
{\large Jacob~Tunesi}$^b$,
{\large  Rachel~M.~Godun}$^b$,
{\large Xavier~Calmet}\footnote{E-mail: x.calmet@sussex.ac.uk}$^a$
}
 \end{center}
\begin{center}
$^a${\sl Department of Physics and Astronomy, \\
University of Sussex, Brighton, BN1 9QH, United Kingdom
}\\
$^b${\sl National Physical Laboratory\\
Hampton Road, Teddington TW11 0LW, United Kingdom}\\

\end{center}
\vspace{2cm}
\begin{abstract}
\noindent
We present a new framework to study the time variation of fundamental constants in a model-independent way. 
Model independence implies more free parameters than assumed in previous studies.
Using data from atomic clocks based on $^{87}$Sr, $^{171}$Yb$^+$ and $^{133}$Cs, we set bounds on parameters controlling the variation of the fine-structure constant, $\alpha$, and the electron-to-proton mass ratio, $\mu$. We consider variations on timescales ranging from a minute to almost a day.  In addition, we use our results to derive some of the tightest limits to date on the parameter space of models of ultralight dark matter and axion-like particles.
\end{abstract}
\vspace{5cm}
\end{titlepage}

\date{\today}

\section{Introduction}
\label{Sec:intro}

The standard model and general relativity are the currently accepted theories of elementary particles and gravity.  
Their predictions are largely controlled by a set of free parameters known as fundamental constants, which are extracted experimentally and assumed independent of time and spatial position.
The underlying origins and potential spacetime variability of the fundamental constants have been rich subjects of investigation, dating back to 
Dirac's large numbers hypothesis~\cite{Dirac:1937ti,Dirac:1938mt}. 

In the years since, dynamical mechanisms, for example from string theory, have been suggested to explain the constants' origins. In some regimes, additional scalar fields imply spacetime variations~\cite{Marciano:1983wy,Kolb:1985sj,Barrow:1987sr,Maeda:1987ku,Barr:1988xw}. 
More generally, many realistic models give rise to variations of fundamental constants (comprehensive reviews may be found in, e.g.,  Refs.~\cite{Uzan:2011,Martins:2017yxk}). 
Models based on quantum field theory,
such as Bekenstein~\cite{Bekenstein} or Barrow~\cite{Barrow:1999is,Barrow:1998df} models 
can describe fundamental-constant variations in terms of low-energy effective interactions of additional scalar fields coupled to the standard model. More recently, variations of the dimensionless constants due to so-called ultralight fields, which could be related to dark matter~\cite{Derevianko14,Stadnik14,Arvanitaki15}, have revived significant attention in this idea (for a recent review see, e.g.,~\cite{NewHorizons}). Investigating the variability of fundamental constants is strongly tied to the foundational assumptions 
and outstanding problems in modern physics.

This work reports on new developments in studying temporal variations of the dimensionless fundamental constants.
In Sec.~\ref{Sec:2}, a generic approach based on effective field theory, applicable particularly to the analysis of temporal variations, is described. The explicit spacetime dependence of a generic dimensionless constant is represented by a series expansion of a scalar field normalized to the energy scale of new physics responsible for the time variation.  A description of scalar-field evolution, including arbitrary damping effects, is given, along with the number of observable parameters. This setup covers a broad range of models describing temporal variations in the literature. It also has significant consequences for the interpretations of measurements, demonstrating that there are more free parameters than typically expected.

Many previous experiments have used the extreme precision of atomic clocks to investigate time-variations of fundamental constants.  Examples include constraints placed on linear-in-time variations~\cite{Rosenband2008, Godun2014, Huntemann2014, Lange21}, oscillations~\cite{Hees2016:DM, Kennedy:2020cavity, BACON:2021DM, Kobayashi:2022vsf, PTB2023} and transients~\cite{Wcislo2018, Roberts2020}.  For cases where the experimental data has been interpreted to place bounds on theoretical parameters such as coupling constants, specific models for the scalar fields and their dynamics have always been assumed.  Our approach in this paper, however, allows constraints on time-variation to be presented in a model-independent way, without assuming the functional form of the scalar field. The model-independent bounds can be interpreted in terms of specific models due to the framework presented in Sec.~\ref{Sec:2}, where we work with the most generic functional form for the scalar field and effective field theory methods.

In Sec.~\ref{Sec:data_analysis}, measurements are presented from atomic clocks ($^{87}$Sr, $^{171}$Yb$^+$ optical and $^{133}$Cs microwave) at the National Physical Laboratory (NPL), assessing the temporal variability (stability) of the fine-structure constant, $\alpha$, and the electron-to-proton mass ratio, $\mu = m_e/m_p$, over a period of about two weeks. The frequency ratios between $^{171}$Yb$^+$, $^{87}$Sr and $^{133}$Cs clock transitions place constraints on oscillations in $\alpha$ and $\mu$ at and beyond the previous state-of-the-art.

Using these data, in particular the $^{171}$Yb$^+$/$^{87}$Sr and $^{87}$Sr/$^{133}$Cs ratios, in Sec.~\ref{Sec:modind_limits} model-independent constraints are placed for the first time on low-dimensional couplings of an ultralight scalar field to matter. In Sec.~\ref{Sec:DM}, new constraints are extracted and compared with previous results for the special cases of ultralight scalar and axion-like dark matter. The prospects of future measurement campaigns are discussed in Sec.~\ref{Sec:conclusions}.
\section{Field theory description of varying constants}
\label{Sec:2}

In this section we introduce our generic framework to describe the spacetime variation of fundamental constants. This framework relies on the concept of effective field theory methods (see e.g. \cite{Burgess:2020tbq}). It is a tool of quantum field theory that has been extensively applied to many areas including particle, nuclear, atomic, condensed matter and gravitational physics. The significance of this approach is that it enables calculations and predictions, which can be tested in experiments, that are generic and universal.

Describing the spacetime variation of a fundamental coupling constant  $g(t,\vec{x})$ can be accomplished by promoting the constant to a series expansion involving a scalar field $\phi(t,\vec{x})$: 
\begin{eqnarray}
\label{gcoupling}
g(t,\vec{x})=g_0+\frac{1}{\Lambda}\phi(t,\vec x)+\ldots \ , 
\end{eqnarray}
where $g_0$ is the spacetime-independent contribution and $\Lambda$ is the high-energy scale relevant to the onset of the physics responsible for the spacetime variation of constants. Coupling $\phi(t,\vec{x})$ to conventional matter otherwise described by a coupling $g_0$ is thus accommodated by replacing $g_0 \rightarrow g(\phi)$.

In this paper, we are primarily interested in time-varying fundamental constants and are thus considering only time-dependent scalar fields. The most generic field equation for a scalar field corresponds to a damped harmonic oscillator. From an effective field theory point of view, this is the field equation that corresponds to dimension-four operators: the kinetic term for the scalar field and potential interactions leading to a decay of the scalar field which can be parametrized by operators of dimensions 3 and 4.  The equation of motion for a damped harmonic oscillator is given by
\begin{equation}
\label{reducedEOM}
\ddot \phi + \Gamma \dot \phi+m^2 \phi=0,
\end{equation}
where $\phi(t)$ is a time-dependent scalar field, $m$ is the mass and $\Gamma$ is a damping factor. 

The solution to the field equation depends on the boundary conditions. 
For oscillatory solutions, the classes of behavior are identified by the relation between $m$ and $\Gamma$:
\begin{align}
\phi(t)= \begin{cases}\phi_{0,1} \exp{\left(r_1 t \right)} +\phi_{0,2} \exp{\left(r_2 t \right)}, \quad
\Gamma^2 - 4 m^2  > 0\; \text{(overdamped)}, \vspace{2mm}\\
\exp{\left(-\frac{ \Gamma t}{2}\right)} (\phi_{0,1} + \phi_{0,2}  t), \quad
\Gamma^2 - 4 m^2  = 0\; \text{(critically damped)},\vspace{2mm}\\ 
\phi_0 \exp{\left(-\frac{ \Gamma t}{2}\right)} \cos \left(\omega_d t - \theta \right), \quad
 \Gamma^2 - 4 m^2  < 0\; \text{(underdamped)},  \end{cases}
\end{align}
where $\phi_0$, $\phi_{0,1}$, and $\phi_{0,2}$ are amplitudes, $\theta$ is a phase and
\begin{align}
&r_{1,2}= -\frac{1}{2} \Gamma \pm \frac{1}{2}  \sqrt{ \Gamma^2- 4 m^2},\\
&\omega_d =   \frac{1}{2} \sqrt{| \Gamma^2- 4 m^2|}.
 \end{align}
 
Standard effective field theory methods can be used to describe the interactions of $\phi$ to conventional matter in a general manner. For example, the interactions of the scalar field with the photon field $A_\mu$ and the electron field $\psi_e$ can be described by the Lagrangian:
\begin{align}  
\label{EMcouplings}
\mathcal{L} =  \left(\kappa\phi\right)^n\left(d^{(n)}_\gamma\tfrac{1}{4} F_{\mu\nu}F^{\mu\nu} - d^{(n)}_{m_e} m_e \bar \psi_e \psi_e \right), 
\end{align}     
for positive integer $n$, where $\kappa = \sqrt{4\pi G} = 1/(\sqrt{2}M_P)$ with $G$ being the Newtonian constant of gravitation, $M_P$ being the reduced Planck mass, and $F_{\mu\nu} = \partial_\mu A_\nu - \partial_\nu A_\mu$. Note that we normalize the interactions to the strength of the gravitational interaction, which is the weakest interaction in nature known to date.  However, the  
 dimensionless couplings $d_\gamma^{(n)}$ and $d_{m_e}^{(n)}$ control, respectively, the strength of interactions between $\phi$ and the photon  and the electron. They may be taken as real numbers, and parameterize the magnitude of the time variation of the fine-structure constant and of the electron mass. The definition used in Eq.~\eqref{EMcouplings} sets the high-energy scale $\Lambda$ from Eq.~\eqref{gcoupling} such that $\kappa^n d_j^{(n)} \equiv g_{0,j}/\Lambda_j^n $.
 In that sense, $\kappa^n d_{j}^{(n)}$ are the combinations of parameters that determine the couplings of the scalar field to the different matter components.

The linear scalar-field coupling ($n=1$) and quadratic scalar-field coupling ($n=2$) are the most relevant as they are the lowest order effective operators and thus their effects are expected to be the strongest. Interactions with $n\geq 3$ are suppressed by more powers of the scale of the physics responsible for the time variation and thus very much suppressed if this is a high-energy scale of the order of the Planck scale, for example.

 Note that there are a number of free parameters that need to be fitted to data:\\ $(\phi_0,\omega_d, \Gamma,\theta,d^{(n)}_{m_e}, d^{(n)}_\gamma)$.
We can use this fully generic approach to calculate the shift of the fine-structure constant $\alpha$ due to the scalar field $\phi(t)$ and obtain in the linear case ($n=1$)
\begin{eqnarray}
\frac{\Delta \alpha}{\alpha}&=&\kappa d^{(1)}_\gamma (\phi(t_2)-\phi(t_1)).
\end{eqnarray}
In the physically relevant underdamped regime, we obtain
\begin{eqnarray}
\frac{\dot \alpha}{\alpha}=-\frac{\kappa d^{(1)}_\gamma \phi_0}{2} \exp{\left(-\frac{\Gamma t}{2}\right)} \bigg(  \Gamma \cos\left( \theta - t \omega_{d}\right) + 2 \omega_{d} \sin\left( \theta - t \omega_{d}\right) \bigg).
\end{eqnarray}
In the quadratic case ($n=2$), we have
\begin{eqnarray}
\frac{\Delta \alpha}{\alpha}&=&\kappa^2 d^{(2)}_\gamma (\phi(t_2)^2-\phi(t_1)^2),
\end{eqnarray}
which leads in the underdamped regime to
\begin{eqnarray}
\frac{\dot \alpha}{\alpha}=-\frac{\kappa^2 d^{(2)}_\gamma\phi_0^2}{2} \exp{\left(-\frac{ \Gamma t}{2}\right)} \bigg(  \Gamma \cos\left( \theta - t \omega_{d}\right) + 2 \omega_{d} \sin\left( \theta - t \omega_{d}\right) \bigg).
\end{eqnarray}

In all cases, we have five independent parameters: the coupling constants to matter $d^{(n)}_\gamma$, an amplitude $\phi_0$, a damping factor $\Gamma$, an oscillation frequency $\omega_{d}$ and a phase $\theta$. Note that they may not all be measurable independently. A measurement of a change of $\alpha$ is only sensitive to the product $d^{(n)}_\gamma \phi^{n}_0$.

Similarly, the coupling of $\phi$ to gluons is controlled by the coupling $d_g^{(n)}$ and that to quarks is controlled by the quark coupling constant $d_{m_f}^{(n)}$. These free parameters control the degree of time variation of the proton mass and of quark masses. We note that the vacuum energy due to gluons, which corresponds to the Quantum Chromodynamics (QCD)-scale $\Lambda_{\rm QCD}$, accounts for approximately 90\% of the nucleon mass $M_N$.  On the other hand, light-quark masses $m_u$, $m_d$ and $m_s$ account for a mere $\sim$10\% of the nucleon masses \cite{Gasser:1982ap}. The proton mass is thus much more sensitive to a time variation of $\Lambda_{\rm QCD}$ than it is to a variation of the light quark masses if all of these parameters vary with time. For more details on the QCD couplings, please see Appendix \ref{Ap1}.

The main motivation for this methodology is that a broad range of models is encompassed by this effective field theory approach. By providing bounds on a time variation of fundamental constants using this formalism, bounds for specific models can easily be obtained and our results can then be interpreted in specific  models, for example: 
\begin{itemize}
\item Quintessence-like models, see, e.g., \cite{Carroll98,Martin08,Dvali:2001dd}: in that case $\Gamma=3H$ where $H$ is the Hubble parameter, but because today $H\sim 1/t_{\rm today}$ (where $t_{\rm today}$ is the age of the universe),  the damping is irrelevant today. Note that for quintessence models, the mass of the scalar field is of order $H\sim 10^{-33}$ eV and there is some tension with torsion pendulum experiments (E\"ot-Wash see, e.g., \cite{Lee:2020zjt}) as these experiments exclude new bosons with masses lighter than $10^{-2}$ eV for $d_j^{(i)}\sim 1$, one needs to consider models with $d_j^{(i)}\ll1$.
\item Ultralight dark matter~\cite{Derevianko14,Stadnik14, Arvanitaki15}: we must assume $\Gamma=0$ as dark matter is stable. As described in detail in Sec.~\ref{Sec:DM}, if the scalar field with mass $m$ accounts for all of dark matter, we can relate $\phi_0$ to the local density of dark matter $\rho_{\rm DM}$: $\phi_0 \approx \sqrt{2\rho_{\rm DM}}/m$. Note that we could have a multi-component dark matter sector in which case the relation between $\phi_0$ and $\rho_{\rm DM}$ doesn't hold. This is further discussed in Sec.~\ref{Sec:DM}.
\item The scalar field could be a generic scalar from some hidden sector~\cite{Calmet:2019frv}. In that case, the amplitude is a free parameter  and $\Gamma \neq 0$ if it can decay today.
\item Kaluza-Klein and moduli models: in models with extra dimensions the sizes of compactified extra dimensions can be described by  scalar fields (the moduli fields). If the size of extra dimensions changes with cosmological time, we could have a time evolution of these scalar fields. In particular, in string theory, all coupling constants are determined by the expectation values of moduli fields.   Coupling constants could thus easily depend on time \cite{Marciano:1983wy,Kolb:1985sj,Barrow:1987sr}. 
\item Dilaton fields, see, e.g., \cite{Campbell:1994bf,polchinski1,polchinski2}: they are similar to moduli, but we expect them to couple universally to matter like gravity does. These models include Brans-Dicke type fields and also scalar fields that are coupled non-minimally to the curvature scalar e.g. $\phi^2 R$, where $R$ is the Ricci scalar.
\item Vacuum evolution models: in some extensions of the Standard Model, the vacuum expectation value of the Higgs boson can evolve with time \cite{Calmet:2017czo,Calmet:2019nfj}. As the vacuum expectation value of the Higgs boson fixes the weak scale, it will lead to a time variation of all fermion masses as well as that of the  electroweak bosons masses. This is typical of inflation-type models.
\item Quantum gravity: quantum gravity predicts that $d_j^{(i)}\sim1$ \cite{Hill:1983xh,Vayonakis:1993nn,Calmet:2009uz,Calmet:2019jyz,Calmet:2020pub,Calmet:2022bin}. Clocks have the capacity of probing the gravitationally generated interactions between very light scalar fields and photons or electrons (i.e. with $d_j^{(i)}\sim 1$) and in that sense they can probe quantum gravity.
\item Test of grand unified theories \cite{Marciano:1983wy,Calmet:2001nu,Calmet:2002ja,Calmet:2002jz,Calmet:2006sc}: in grand unified models, time shifts in $\alpha$ and the strong coupling constant $\alpha_s$ (equivalently $\Lambda_{\rm QCD}$) are related. If time variations of both $\alpha$ and $\mu$ were observed, predictions of grand unified theories could be directly probed.  The same observation applies to shifts in lepton and quark masses. Because the relations between quark and lepton masses are very model dependent, clocks could help to determine the correct unification theory using very low energy data without the need to produce super massive particles.
\end{itemize} 
 
 We thus see that our generic theoretical framework enables us to study a very wide variety of models and also to test different scenarios of ultra-high-energy physics with very low-energy experiments. This is obviously only a subset of the models that can be studied with these methods. Note that our approach enables us to describe both fundamental variations of constants as originally envisaged by Dirac, but also the effective time evolution of constants where the variation is due to interactions with additional fields, as in the case of, for example, ultralight dark matter. We point out that this framework can also be applied to massive spin-1 and spin-2 fields with only minor modifications in terms of the way these higher spin fields couple to matter.

 It should also be emphasized that while we looked at a damped oscillator model, there are important other classes of models that could have been considered. Depending on the boundary conditions, other solutions are possible, e.g., soliton models, transient phenomena, cosmic strings, domain wall, kink solution, etc. Great care needs be taken when interpreting data as new physics signals could easily get lost when a specific functional form of the signal is assumed.

Note that for infinitesimal time differences between two measurements of the fundamental constants, we can consider the first-order shifts
\begin{eqnarray}
    \label{variations}
\frac{\delta\alpha}{\alpha} &\equiv& d_\gamma^{(n)}(\kappa\phi)^n, \\ \nonumber \frac{\delta m_f}{m_f} &\equiv& d_{m_f}^{(n)}(\kappa\phi)^n, \\ \nonumber  \frac{\delta \Lambda_{\text{QCD}}}{\Lambda_{\text{QCD}}} &\equiv& d_g^{(n)}(\kappa\phi)^n,
\end{eqnarray}
with $f = e, u, d$. 

For the data analysis performed in the next section, it is useful to consider fractional frequency shifts in atomic clocks $\delta\nu/\nu$, where $\nu$ is the frequency of an atomic transition. Since these fractional frequencies depend on at least one fundamental constant, new physics in the form of $g(\phi)$ altering spectral widths could be imparted on laboratory measurements. As observables involve comparisons with a reference, in this context fractional frequency ratios are considered. These ratios are related to a linear combination of relative fundamental constant variations,
\begin{equation}
\label{eq:df=KdX}
\frac{\delta (\nu_1/\nu_2)}{\nu_1/\nu_2} = \sum_{g} (K_{g_1} - K_{g_2})\frac{\delta g}{g},
\end{equation}
where $K_{g_i}$ is a sensitivity coefficient particular to the system of interest~\cite{Flambaum:1999VFCs-A,Flambaum:1999VFCs-B,Flambaum:2009VFCs}. The frequency of optical and microwave transitions can be expressed as
\begin{align}
\label{freqs}
&\nu_{\rm opt} = A\, F_{\rm opt}(\alpha)\, cR_{\infty}, \nonumber\\
&\nu_{\rm MW} = B\,\alpha^2 F_{\rm MW}(\alpha)\, g_{N}\, (m_e/m_p) \, cR_{\infty},
\end{align}
where $A$ and $B$ are constants specific to a given transition, $F_{\rm opt}(\alpha)$ and $F_{\rm MW}(\alpha)$ are relativistic corrections and $g_{N}$ is the nuclear $g$-factor, $c$ is the speed of light and $R_{\infty}$ is the Rydberg constant. This work focuses on optical-to-optical Yb$^+$/Sr and optical-to-microwave Sr/Cs ratios. The former are sensitive to $\alpha$ variations whereas the latter are sensitive to $\alpha, \mu$ and $g_N$ variations.\footnote{Very recent works have claimed optical-to-optical ratios are also sensitive to oscillations of nuclear charge radii~\cite{radii1,radii2}.} We now turn out attention to the data analysis.

\section{Data and analysis framework}
\label{Sec:data_analysis}
\subsection{Experimental Setup}
In this work we analyze frequency ratio data produced by atomic clocks based on neutral strontium atoms in a lattice trap (Sr) \cite{Hobson20}, a singly-charged ytterbium ion in a Paul trap (Yb$^{+}$) \cite{Baynham2018Ytterbium}, and neutral cesium atoms launched in a fountain configuration (Cs) \cite{Szymaniec2016}. The properties of the atomic and ionic energy transitions are summarized below in Table \ref{tab:freq info}.

\begin{table}[ht]
\centering
\begin{tabular}[t]{|c|c|c|c|c|}
\hline\hline
Species&Transition&$K_{cR_{\infty}}+K_{\alpha}$&$K_{\mu}$&$K_{q}$\\
\hline
$^{87}$Sr & $^{1}$S$_{0}- ^{3}$P$_{0}$ & $2+0.06$ & 0.00 & 0.00\\
$^{171}$Yb$^{+}$ & $^{2}$S$_{1/2}- ^{2}$F$_{7/2}$ & $2-5.95$ & 0.00 & 0.00\\
$^{133}$Cs & $^{2}$S$_{1/2}(F=3 \rightarrow F=4)$ & $2+2.83$ & 1.00 & 0.07\\
\hline\hline
\end{tabular}
\caption{Summary of the atomic and ionic energy transitions used to produce high-stability frequency data. $K_X$ values are taken from references \cite{Flambaum:1999VFCs-B, Dzuba2008, berengut, Angstmann2004, FlambaumTedesco} and given to two decimal places.}
\label{tab:freq info}
\end{table}

\begin{figure}[ht]
\centering
    \includegraphics[width=0.8\textwidth]{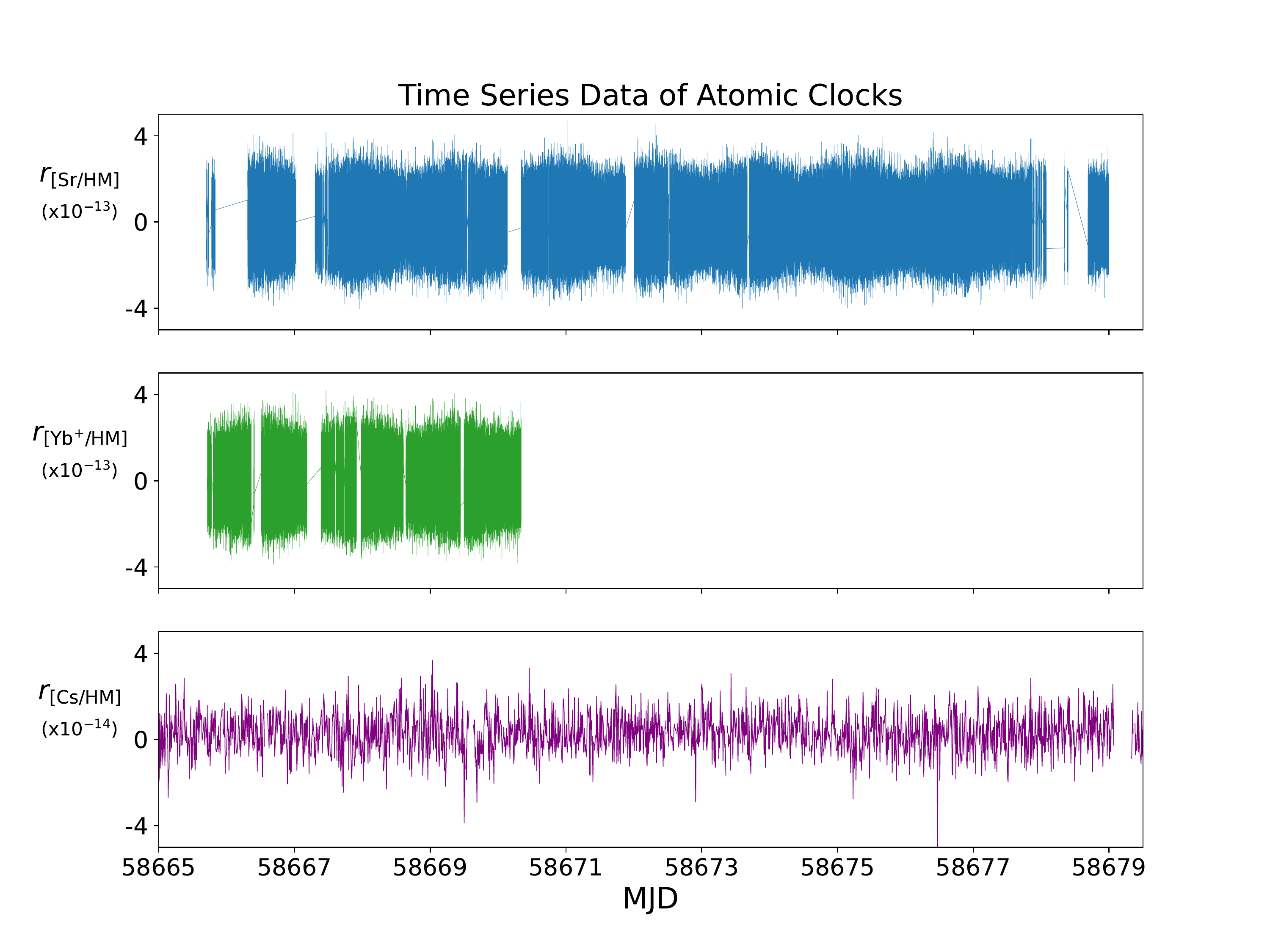}
    \caption{Top and middle: Fractional frequency ratios of Sr/HM and Yb$^{+}$/HM data counted by NPL's femtosecond frequency comb, with data averaged over 1~s intervals. Bottom: Fractional frequency ratio of Cs/HM data produced by NPL's cesium fountain, NPL-CsF2, with data pre-processed over 600~s intervals. Data were collected between $1^{\text{st}}$-$14^{\text{th}}$ July 2019 (MJD 58665 - 58679).}
    \label{fig:time_series}
\end{figure}

The Sr, Yb$^{+}$, and Cs clock frequencies are all measured relative to an active hydrogen maser (HM). For the optical clock frequencies from Sr and Yb$^{+}$, this
measurement is made via a frequency comb, referenced to the 10~MHz output of the maser.  The microwave clock frequency from Cs, however, can be measured directly against the maser. As all the measurements are made  during the same observation window, the frequency ratios Yb$^{+}$/Sr, Yb$^{+}$/Cs and Sr/Cs can be calculated in post-processing and are independent of the maser frequency.  The achieved stabilities in the Yb$^{+}$/Sr, Yb$^{+}$/Cs and Sr/Cs frequency ratios are all close to the quantum projection noise limits, determined by the number of atoms interrogated, the clock cycle times and the quality of the local master oscillator~\cite{Schioppo22}.  Fig.~\ref{fig:time_series} displays the time series of data used in this work, plotted as fractional frequency ratios:
\begin{eqnarray}\label{ffrdef}
r_{[i/j]}= \frac{(\nu_{i}/\nu_{j} - R^{*}_{ij})}{R^{*}_{ij}} \, ,
\end{eqnarray}
with reference ratios, $R^{*}_{ij}=\nu^{*}_{i}/\nu^{*}_{j}$, defined using the reference frequency values: $\nu^{*}_{\text{Sr}}=429,228,004,229,872.99$~Hz, $\nu^{*}_{\text{Yb}^{+}}=642,121,496,772,645.12$~Hz, $\nu^{*}_{\text{Cs}} = 9,192,631,770$~Hz \cite{BIPMfreqs}, and with the reference value $\nu^{*}_{\text{HM}}$ taken to be 10 MHz. The Sr and Cs data were available between $1^{\text{st}}$- $14^{\text{th}}$ July 2019, with total uptimes of 73\% and 93\% respectively; Yb$^{+}$ data were available between $1^{\text{st}}$- $6^{\text{th}}$ July 2019, with a total uptime of 76\%.

\subsection{Experimental Results: Frequency Ratio Instability}

In the presence of general noise processes, the mean and standard deviation of data sets are not guaranteed to converge as the number of samples increases. It is therefore common practice to estimate the spread of $r_{[i/j]}$ over different averaging times, $\tau$, using the Allan Deviation, $\sigma_{r}(\tau)$, and its extensions. These are more informative estimators of spread, as they converge for data sets exhibiting the most common kinds of non-stationary statistics \cite{NISTNote1337,NISTbooklet}. Specifically in this work, we characterise the instability of our data using the Modified Allan Deviation (MDEV). The MDEV is given by the square root of the Modified Allan Variance, which is defined for a data set of $M$ measurements, $y_k$, averaged over averaging time, $\tau$, as
\begin{eqnarray}
\label{mdevdef}
\sigma^{2}_{y}(\tau) = \frac{\tau^{4}_{0}}{2\tau^4(M-3m+2)} \sum_{j=1}^{M-3m+2} \left\{ \sum_{i=j}^{j+m-1} \left( \sum_{k=i}^{i+m-1} [ y_{k+m} - y_{k}] \right) \right\}^2 \, ,
\end{eqnarray} 
where $m$ is the averaging factor, such that $\tau = m \tau_{0}$, and $\tau_0$ is the original sampling time interval of the data points \cite{NISTbooklet}. As our frequency data were measured on a frequency counter configured in a $\Lambda$-counting mode, the MDEV was the appropriate statistic to characterise the instability \cite{Rubiola2005, Dawkins2007, Benkler2015}. To directly compare instability estimates from different types of Allan deviation, the estimators may be converted between each other using the relations documented in \cite{Rubiola2023}.

\begin{figure}[ht]
  \centering
    \includegraphics[width=0.8\textwidth]{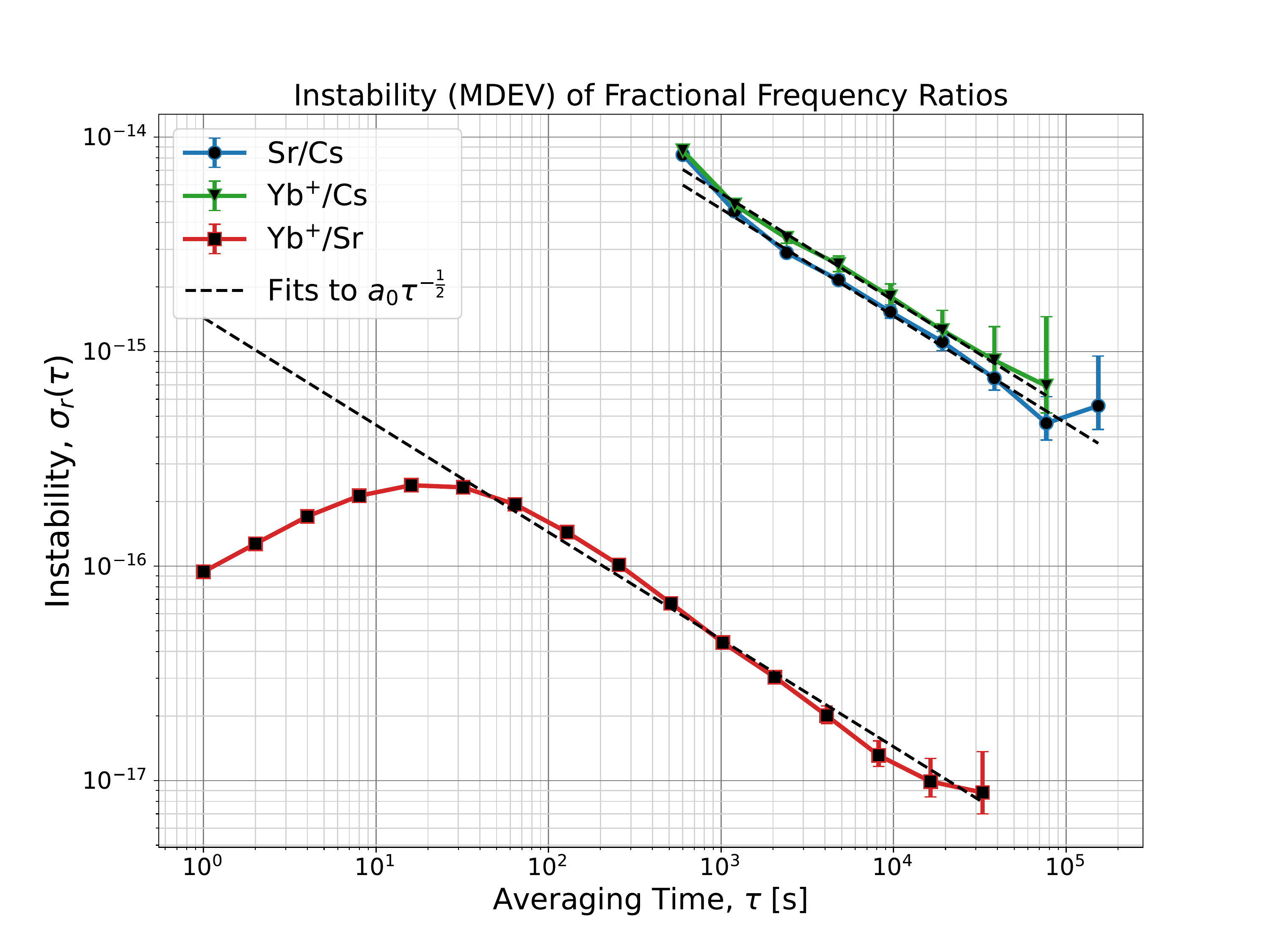}
    \caption{Characterisation of each fractional frequency ratio's instability using the Modified Allan Deviation, plotted at octave intervals of averaging time.}
    \label{fig:raw_MDEVs}
\end{figure}

Fig.~\ref{fig:raw_MDEVs} displays the values of $\sigma_{r}(\tau)$ calculated at octave intervals of averaging time. For 60~s~$\leq\tau\leq$~30~000~s, the $\sigma_{r[\text{Yb}^{+}\text{/Sr}]}$ curve is well approximated by power-law behavior of $\sigma_{r[\text{Yb}^{+}\text{/Sr}]}(\tau)\approx1.4\times 10^{-15}/\sqrt{\tau /{\rm s}}$. This indicates that the data are well described on these timescales by a white frequency modulation (WFM) noise process with $h_0 \approx 8.3  \times  10^{-30}$~Hz$^{-1}$, where $h_0$ is a constant value of power spectral density~\cite{NISTNote1337, NISTbooklet}. This is not true for $\tau<60$~s, because this is shorter than the time constant for steering the Yb$^{+}$ clock laser onto the atomic transition frequency.  At these shorter timescales, correlated noise from the clock laser dominates the instability. For averaging times $600$~s $\leq \tau \leq 80~000$~s, the instabilities of Sr/Cs and Yb$^{+}$/Cs are well approximated by $\sigma_{r[\text{Sr/Cs}]} \approx \sigma_{r[\text{Yb}^{+}\text{/Cs}]} \approx 1.6\times 10^{-13}/\sqrt{\tau / {\rm s}}$, corresponding to WFM noise with $h_0\approx 1.0 \times 10^{-25}$~Hz$^{-1}$. The small difference in instability between the Yb$^{+}$/Cs and Sr/Cs ratios can be attributed to the fact that the two sets of data span different time periods with different uptimes. While $\sigma_{r[\text{Sr/Cs}]}$ appears to increase at the highest averaging time, we attribute this to error introduced by the routine used to calculate MDEVs, which interpolates gaps in the data record: when downtime is dominated by a small number of large gaps in the data record (as is the case here), this interpolation can introduce spurious drifts and inflate the instability estimate at the largest averaging times \cite{RileyGaps, Stable32}. Therefore we do not consider averaging times above $\tau = 80~000$~s.

Some publications \cite{Kennedy:2020cavity} have attempted to constrain variations in fundamental constants using frequency data from hydrogen masers (HM), reasoning that $\nu_{\text{HM}}$ shares the $K_{X}$ sensitivities of the $^{2}$S$_{1/2}$ (F=0 $-$ F=1) hyperfine transition in atomic hydrogen to which the maser cavity is tuned. We do not follow this approach in this work, as we cannot confirm over which timescales this reasoning holds true in our commercial maser system; the maser wall shift, the resonant cavity, the voltage-controlled crystal oscillator, etc. all contribute to the instability of $\nu_{\text{HM}}$ over certain timescales and introduce sensitivity to additional variables, e.g. temperature \cite{karshenboim2000}. Consequently, we do not use optical-to-maser ratio data (Sr/HM or Yb$^{+}$/HM) directly in this work to place constraints on the variations in fundamental constants. Instead, we use optical-to-cesium ratio data, despite the higher WFM instability in the data sets, as we are confident in the sensitivities of $\nu_{\text{Cs}}$ to variations in the fundamental constants.

\begin{figure}[ht]
  \centering
  \includegraphics[width=0.8\textwidth]{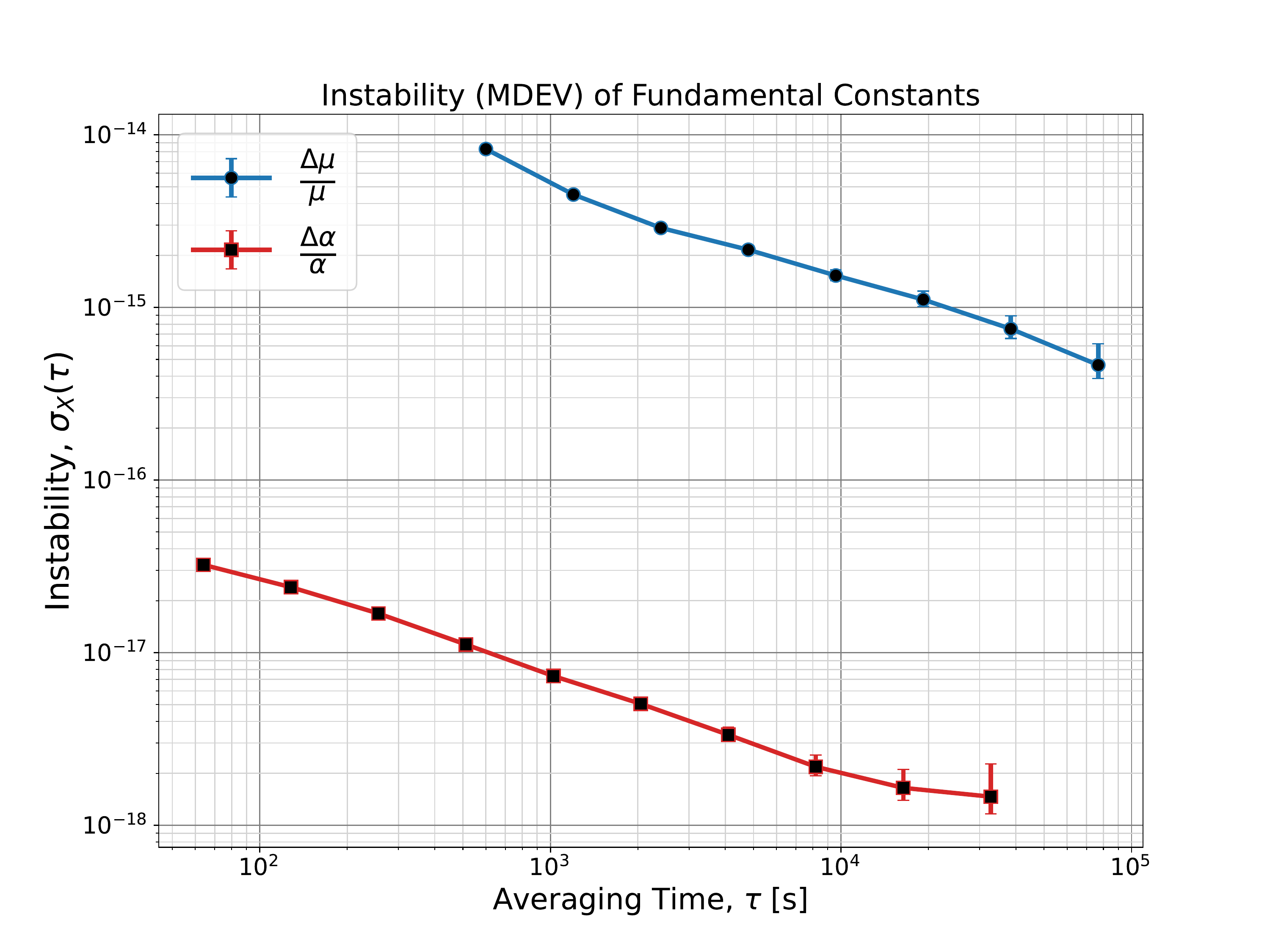}
    \caption{Instability estimates for fractional changes in $\alpha$ and $\mu$, where $\sigma_{X}(\tau) \equiv \sigma_{(\Delta X / X)}(\tau) = \sigma_{r}(\tau) / K^{r}_X$ on timescales over which the instability is dominated by the behavior of the atomic transition frequency.}
    \label{fig:alphamu_MDEVs}
\end{figure}

On timescales for which the instability of the frequency ratio data is dominated by the behavior of the atomic transitions, we place constraints on fundamental constants using Eq.~\eqref{eq:df=KdX} and $K_{X}$ values taken from Table~\ref{tab:freq info}. Due to the negligible sizes of $\Delta K_{\mu}^{\text{Yb$^{+}$/Sr}}$ and $\Delta K_{q}^{\text{Yb$^{+}$/Sr}}$, we assume that the instability in fractional changes in $\alpha$ to leading order must be less than $\sigma_{r[\text{Yb$^{+}$/Sr}]}$ scaled by the magnitude of the sensitivity $\Delta K_{\alpha}^{\text{Yb$^{+}$/Sr}} = | K_{\alpha}^{\text{Yb$^{+}$}}~-~K_{\alpha}^{\text{Sr}}| = 6.01$. Thus, we constrain the instability of fractional changes in $\alpha$ to be $\sigma_{(\Delta\alpha/\alpha)} = \sigma_{r[\text{Yb$^{+}$/Sr}]} / 6.01 \leq 2.3\times 10^{-16}/\sqrt{\tau / {\rm s}}$ on timescales of 60~s $\leq \tau \leq 30~000$~s. With $\sigma_{(\Delta\alpha/\alpha)}$ constrained to two orders of magnitude below the noise level of Sr/Cs and Yb$^{+}$/Cs, we make the further assumption that any remaining instability in $r_{[\rm Sr/Cs]}$ would be dominated by $\Delta \mu / \mu$ rather than $\Delta m_{q} /m_{q}$, due to the small size of $\Delta K_{q}^{\text{Sr/Cs}} = |K_{q}^{\text{Sr}}~-~K_{q}^{\text{Cs}}| \approx 0.07$ compared with $\Delta K_{\mu}^{\text{Sr/Cs}} = |K_{\mu}^{\text{Sr}}~-~K_{\mu}^{\text{Cs}}| = 1.00$. Under this assumption, to leading order we may similarly constrain the instability of fractional changes in $\mu$ to be no greater than $\sigma_{r[\text{Sr/Cs}]}$ scaled by the magnitude of the sensitivity $\Delta K_{\mu}^{\text{Sr/Cs}} = 1.00$. Thus, we constrain the instability of fractional changes in $\mu$ to be $\sigma_{(\Delta\mu/\mu)} =$ $\sigma_{r[\text{Sr/Cs}]} / 1.00$ $\leq 1.6\times 10^{-13}/\sqrt{\tau / {\rm s}}$ on timescales of $600$~s~$\leq \tau \leq 80~000$~s. These constraints on the instability of fractional changes in $\alpha$ and $\mu$ as a function of averaging times are shown in Fig.~\ref{fig:alphamu_MDEVs} and summarized in Table~\ref{tab:mdev constraints}.

These estimates of fractional variations in frequency ratios, and hence $\alpha$ and $\mu$, on different timescales make no assumptions about the functional form of the variations.  These results can be translated into model-independent limits, which will be discussed in Sec. \ref{Sec:modind_limits}.

\begin{table}[ht]
\centering
\begin{tabular}[ht]{|c|c|}
\hline\hline
Instability Constraint & Parameter space\\
\hline 
\multirow{2}{*}{$\sigma_{(\Delta \alpha / \alpha)}(\tau) \le 2.3\times 10^{-16}/\sqrt{\tau /{\rm s}}$} & \multirow{2}{*}{$60$~s $\le \tau \le 30~000$~s} \\
 & \\
\multirow{2}{*}{$\sigma_{(\Delta \mu / \mu)}(\tau) \le 1.6 \times 10^{-13}/\sqrt{\tau /{\rm s}}$} & \multirow{2}{*}{$600$~s $\le \tau \le 80~000$~s}  \\
 & \\
\hline\hline
\end{tabular}
\caption{Summary of constraints on the instability of fractional changes in $\alpha$ and $\mu$ produced in this paper, expressed as Modified Allan Deviations, $\sigma_{(\Delta X /X)}(\tau)$.}
\label{tab:mdev constraints}
\end{table}

\subsection{Experimental Results: Sinusoidal Oscillations}
If one chooses to focus specifically on oscillatory time variations of fundamental constants, these can be generically described by a damped harmonic oscillator, given in Eq.~\eqref{reducedEOM}. Constraints on damped oscillatory signals could be obtained for a range of oscillation frequencies, $f$, by finding the best fit amplitude of each oscillation frequency, $A(f)$, and the best fit to the damping factor, $\Gamma$. However, it was decided as a first stage of analysis to fit undamped oscillations $(\Gamma=0)$ to reduce the number of parameters that require fitting. If any significant signals or features were detected as a result of fitting to undamped oscillations, reasonable values of $\Gamma$ could be inferred from the linewidths of any peaks, and further analysis could be performed to fit for $\Gamma$ and constrain damped oscillations to the data. In the case that significant features were observed, we implemented a routine to fit the data to oscillations weighted by an envelope function, $Z(t)$, which for $Z(t) = \exp(-\Gamma t / 2)$ would model underdamped oscillations \cite{Bretthorst2006}\footnote{Since our longest data set had a length of $T=14$ days and our minimum sampling time was $\Delta t$ = 60 s, the approximate range of $\Gamma$ detectable with our data was between $\Gamma_{\rm min} \sim 1/T \approx 8.3 \times 10^{-7}$~Hz and $\Gamma_{\rm max} \sim 1 / \Delta t \approx 1.7  \times 10^{-2}$~Hz. The range of detectable values for $\Gamma$ corresponds to a range of lifetimes between 60 s to $1.2\times 10^6$ s, using $\tau^{*}=\Gamma^{-1}$. It is interesting to compare this range to lifetimes corresponding to known forces of nature. Typical lifetimes in Quantum Electrodynamics are of the order of $10^{-20}$~s to $10^{-16}$~s. For the weak interactions, one finds $10^{-13}$~s to $10^{3}$~s while for the strong force one finds $10^{-23}$~s to $10^{-20}$~s. Thus the range of detectable lifetimes in this work partly overlaps with lifetimes typical of the weak force.}. During testing it was observed that this routine did not perform better than standard periodograms for detecting damped oscillations in the parameter space of interest, so it was decided to use a standard periodogram.

Similar to the approach taken in recent works~\cite{Leefer:2015DM, Kennedy:2020cavity, BACON:2021DM, PTB2023}, we constrain the magnitude of undamped oscillations in our data by estimating the power spectral density of the fractional frequency ratios, $S_{r}(f)$, via the Lomb-Scargle periodogram (LSP) \cite{lomb1976, scargle1982}. The LSP is an estimator of $S_{r}(f)$ for time series that suffer from irregular sampling or data gaps due to experiment downtime \cite{VanderPlas2018}. Calculating the LSP is equivalent to performing linear least-squares fits of a data set to the amplitude of sinusoids at a range of frequencies; it allows algorithms in the style of a fast Fourier transform to be used on time series with incomplete or irregular sampling, without having to account for data gaps by deconvolving the time series with composite window functions \cite{scargle1982, Munteanu2016}.

Power spectral density estimates for the fractional frequency ratios, $S_{r}(f)$, were calculated using the implementation of the LSP \cite{lomb1976, scargle1982} provided in the Astropy Python package \cite{VanderPlas2018}. The Nyquist frequency, $f_{\text{Ny}} = (2\Delta t_{\text{sample}})^{-1}$, was chosen as the bandwidth upper limit for ease of comparison with previous works, and any future works that may have 100\% data uptime. However, $S_{r}(f)$ is not guaranteed to be free of aliasing for $f<f_{\text{Ny}}$ when signals are irregularly sampled, as gaps in data sampling introduce spectral leakage below $f_{\text{Ny}}$ \cite{lomb1976}. The resolution of the frequency grid was not tuned beyond using the default oversampling factor of $n_0=5$ \cite{VanderPlas2018}. The fidelity of LSP in detecting oscillations in noisy data was validated by injecting sinusoidal signals into data sets with similar noise statistics as those of the observed data.
\begin{figure}[ht]
  \centering
    \includegraphics[width=0.8\textwidth]{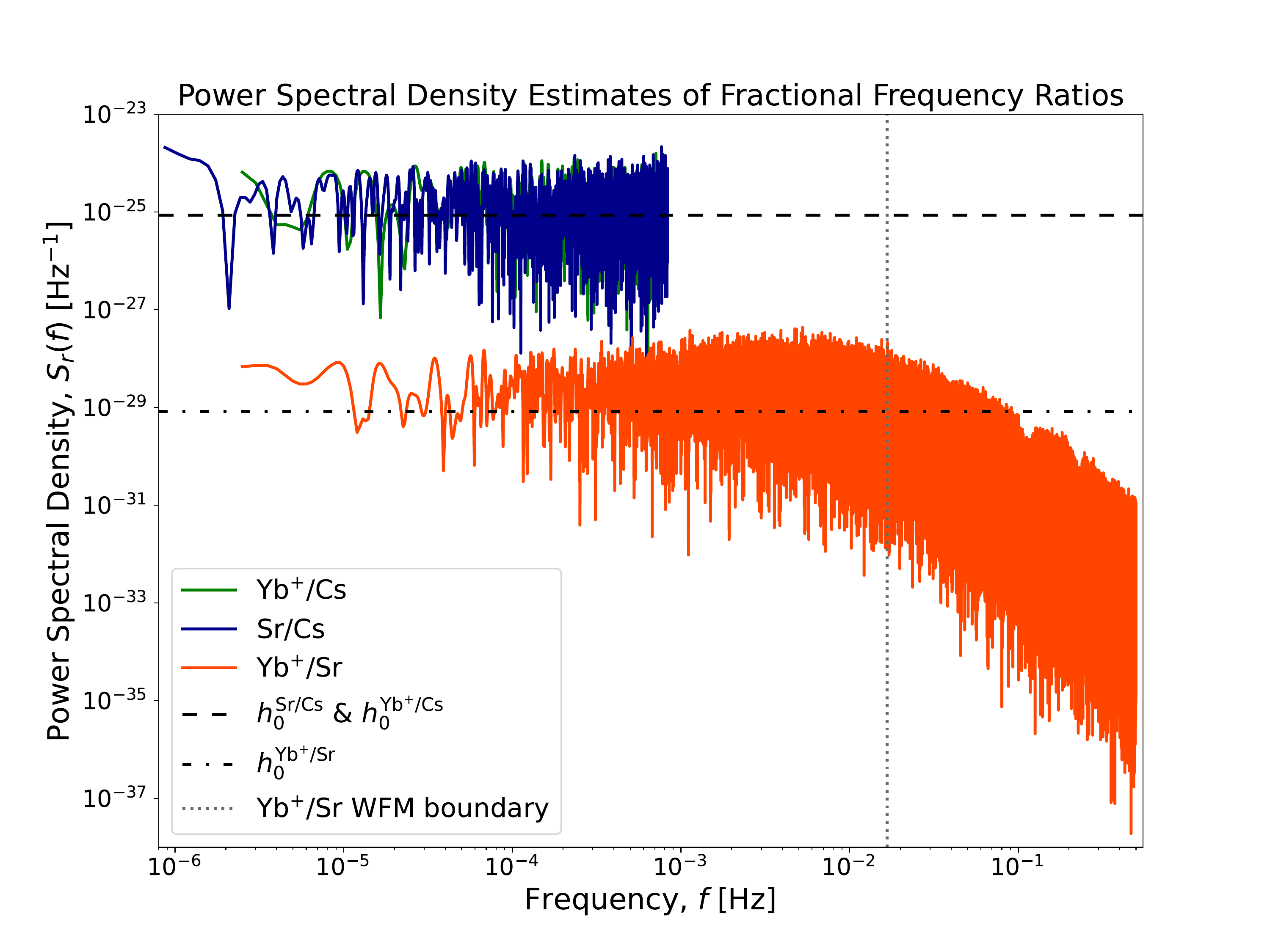}
    \caption{Power spectral densities of fractional frequency ratios, estimated with the Lomb-Scargle periodogram. The levels of white frequency modulation (WFM) noise, $h_0^i$, are taken from the best fits to the $\sigma_{r}(\tau)$ curves presented in Fig.~\ref{fig:raw_MDEVs}.}
    \label{fig:raw_LSP}
    \end{figure}
\begin{figure}[ht]
\centering
    \includegraphics[width=0.8\textwidth]{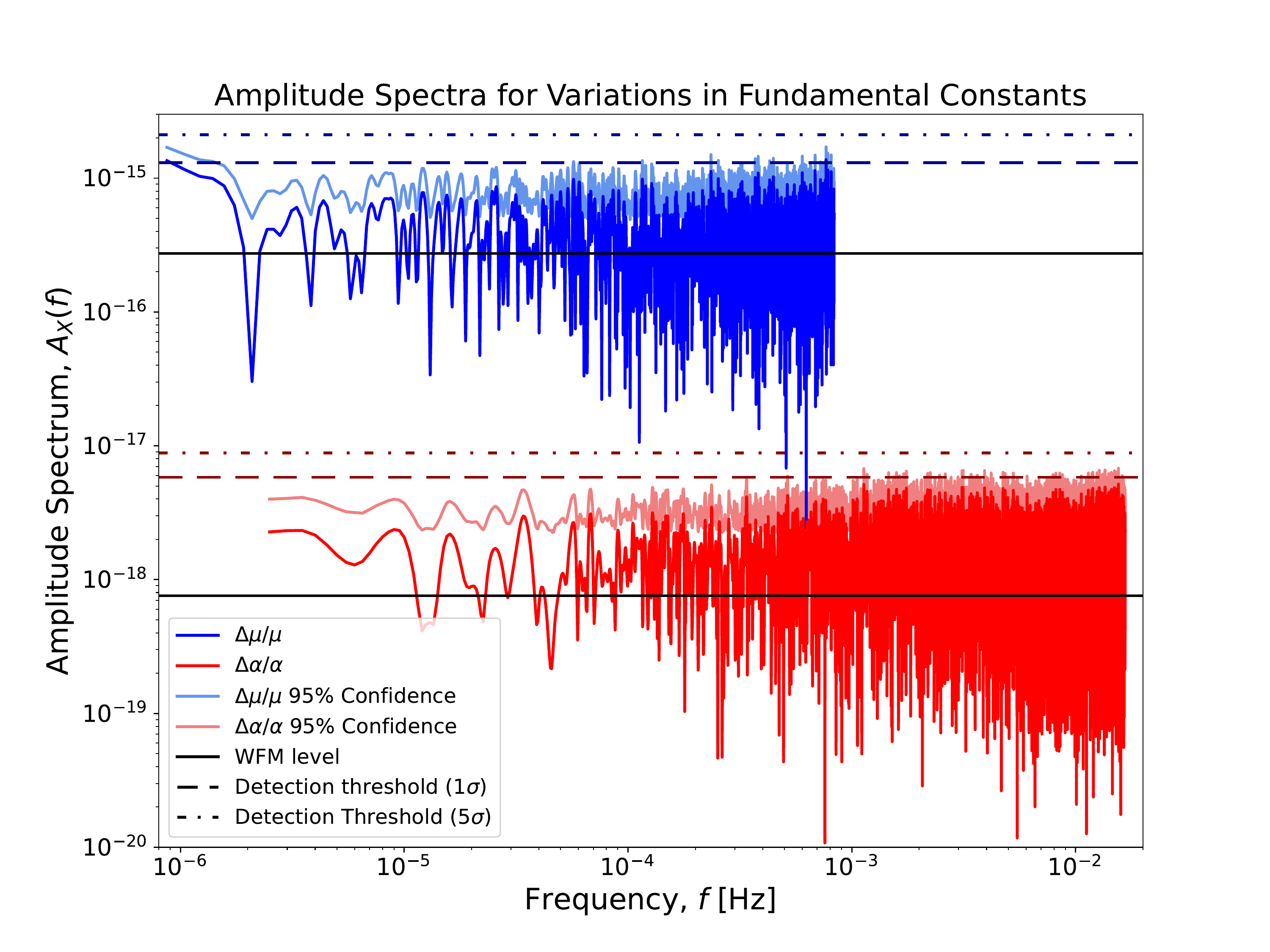}
    \caption{Fractional amplitude spectra for sinusoidal oscillations in $\alpha$ and $\mu$. The solid horizontal lines indicate the estimated levels of white frequency modulation (WFM) noise of the spectra, and the broken horizontal lines represent the estimated false alarm levels at the equivalent of $1 \sigma$~($p \leq 0.32$) and $5 \sigma$~($p \leq 3.5\times 10^{-7}$) significance.}
    \label{fig:alphamu_spectra}
\end{figure}

Fig.~\ref{fig:raw_LSP} shows $S_{r}(f)$ for each frequency ratio, with black dashed lines indicating the estimated noise levels, $h_0$, calculated from values of $\sigma_{r}(\tau)$ \cite{Rubiola2023}. Here it can be seen again that while $r_{[\text{Sr/Cs}]}$ and $r_{[\text{Yb}^{+}\text{/Cs}]}$ are well described by WFM noise, this is not true of $r_{[\text{Yb}^{+}\text{/Sr}]}$ for $f \geq (60\text{ s})^{-1} = 1.67 \times 10^{-2}$~Hz, where more complex power-law behaviour can be seen: the action of the servos steering the probe light frequency onto the atomic/ionic transition frequencies leads to an approximate $S_{r}(f)\propto f^{-1.5}$ behavior. Limits on oscillations in $\Delta\alpha/\alpha$ in the frequency range that exhibits WFM noise can be formed by integrating $S_{r[\text{Yb}^{+}\text{/Sr}]}(f)$ over the nominal frequency bin width ($\delta f_{[\text{Yb}^{+}\text{/Sr}]} = 1/T_{[\text{Yb}^{+}\text{/Sr}]} = 2.5 \times 10^{-6}$~Hz) to obtain the total power of each bin, then taking the square-root to obtain the one-sided\footnote{The Astropy LSP returns only the $f>0$ half of the two-sided amplitude spectrum \cite{VanderPlas2018} which must be multiplied by 4 to obtain the one-sided amplitude spectrum.} fractional amplitude spectrum, $A_{r[\text{Yb}^{+}\text{/Sr}]}(f)$.

Dividing this spectrum by the magnitude of $\Delta K_{\alpha}^{\text{Yb}^{+}\text{/Sr}}$ yields the fractional amplitude spectrum of $\Delta \alpha / \alpha$ oscillations: $A_{\alpha}(f) = A_{r[\text{Yb}^{+}\text{/Sr}]}(f) / 6.01 \approx 7.6 \times 10^{-19}$. Similarly, we constrain oscillations in $\Delta\mu/\mu$ using $S_{r[\text{Sr/Cs}]}(f)$ ($\delta f_{[\text{Sr/Cs}]}=1/T_{[\text{Sr/Cs}]} = 8.7\times 10^{-7}$~Hz) resulting in a fractional amplitude spectrum of $\mu$ as $A_{\mu}(f) \approx A_{[\text{Sr/Cs}]}(f) / 1.00$ $\approx 2.7 \times 10^{-16}$. Both fractional amplitude spectra $A_{\alpha}(f)$ and $A_{\mu}(f)$ are shown in Fig.~\ref{fig:alphamu_spectra} for the frequency ranges over which they may be used to constrain oscillations in $\alpha$ and $\mu$.

Constraints on oscillations in $\alpha$ and $\mu$ can be extracted from the fractional amplitude spectra $A_{\alpha}(f)$ and $A_{\mu}(f)$ by considering the statistics of the LSP estimator. In the absence of prominent peaks, some authors \cite{BACON:2021DM} have placed constraints at the observed power for each frequency bin of the spectra. Following  the approach taken in \cite{Leefer:2015DM,Hees2016:DM, Kennedy:2020cavity, PTB2023} and others, confidence intervals on the constraints at each frequency bin were calculated by simulating a large number of control spectra with equivalent noise statistics to the observed spectra, then using the empirical cumulative distribution function of the simulated power in each frequency bin to estimate confidence intervals, in this case, at 95\% confidence. Whilst we follow this approach to calculate 95\% confidence intervals on our power estimates, using these confidence intervals as ``limits'' has the disadvantage that the bounds placed on a WFM process could differ across neighboring frequency bins by an order of magnitude or more, and would likely fluctuate between repeat experiments even if the noise level in the experiment remained constant.

We believe a more appropriate and reproducible bound for a WFM spectrum is one that is constant across all frequencies: under the null hypothesis that the power differences across frequency bins are merely the result of fluctuations due to a WFM noise process, we produce a global bound based on an estimate of the white noise level, with exclusion limits estimated with false alarm levels, $A_{X}(p \leq p_{0})$. These false alarm levels represent the value of $A_{X}(f)$ that would be exceeded with a probability of no more than $p_0$ across all frequencies in the case of white noise only. While analytic expressions exist for the distribution of $S_{r}(f)$ for regularly sampled data (assuming uncorrelated Gaussian errors) \cite{scargle1982, groth1975}, in this work we employ computational methods, as recommended in \cite{VanderPlas2018}.

Simulating data sets with the same noise and data gaps as the observed spectra, we use the bootstrap method \cite{ivezic2014} to estimate bounds at the 68\% significance level of $A_{\alpha}(p \leq 0.32)=5.6\times 10^{-18}$ and $A_{\mu}(p \leq 0.32)=1.3\times 10^{-15}$. More appropriate to particle physics would be the equivalents of a $5\sigma$ significance level, which were estimated using the Baluev method \cite{Baluev2008}, which yield $A_{\alpha}(p \leq 3.5 \times 10^{-7}) \approx 8.9\times 10^{-18}$ and $A_{\mu}(p \leq 3.5\times 10^{-7}) \approx 2.1\times 10^{-15}$ \footnote{The Baluev method was used in favor of the bootstrap method to estimate the $p \leq 3.5\times 10^{-7}$ false alarm levels, as a robust bootstrap estimate for this level would require $> 2.8\times 10^{7}$ simulated spectra for each data set. This large increase in computation time would likely produce 5$\sigma$ false alarm levels that were only slightly stricter: the Baluev method overestimates the false alarm level by design, but agrees reasonably well with the bootstrap method for observation windows with little structure, such as for our data sets \cite{VanderPlas2018}.}. As shown in Fig.~\ref{fig:alphamu_spectra}, all spectral peaks fall well below this fractional amplitude, though it should be admitted that this is a fairly strict detection threshold. Though global bounds across the entire frequency domain of the LSP estimate are more conservative than bounds one could achieve by setting constraints on individual frequency bins, for the reasons outlined above, we believe them to be better motivated and more legitimate for processes that appear to be predominantly WFM noise.

With only two peaks slightly exceeding the $1\sigma$ false alarm level for $A_{\mu}(f)$, and no spectral peaks exceeding the $5\sigma$ false alarm levels for either $A_{\mu}(f)$ or $A_{\alpha}(f)$, we have high confidence that all peaks in $A_{\alpha}(f)$ and $A_{\mu}(f)$ are consistent with instabilities due to WFM noise processes. Therefore, based on the estimated WFM noise levels of fractional frequency ratio data, we place constraints on sinusoidal oscillations in $\alpha$ at the $1\sigma$ significance level of $A_{\alpha}(f) \le 5.6\times 10^{-18}$ and at the $5\sigma$ significance level of $A_{\alpha}(f) \lesssim 8.9\times10^{-18}$, over the frequency range $2.5 \times 10^{-6}$~Hz $\le f \le 1.7\times 10^{-2}$~Hz. We also place constraints on sinusoidal oscillations in $\mu$ at the $1\sigma$ significance level of $A_{\mu}(f) \le 1.3\times 10^{-15}$ and at the $5\sigma$ significance level of $A_{\mu}(f) \lesssim 2.1\times 10^{-15}$, over the frequency range $8.7\times 10^{-7}$~Hz $\le f \le 8.3\times 10^{-4}$~Hz.  A summary of these $1\sigma$ and $5\sigma$ constraints is shown in Table~\ref{tab:oscillation constraints}.

\begin{table}[ht]
\centering
\begin{tabular}[t]{|c|c|c|}
\hline\hline
Amplitude Constraint & Amplitude Constraint & \multirow{2}{*}{Parameter space} \\
($1 \sigma$ significance) & ($5 \sigma$ significance) & \\
\hline 
\multirow{2}{*}{$A_{\alpha}(f) \le 5.6\times 10^{-18}$} & \multirow{2}{*}{$A_{\alpha}(f) \lesssim 8.9\times 10^{-18}$} & \multirow{2}{*}{$2.5\times 10^{-6}$~Hz $\le f \le 1.7\times 10^{-2}$~Hz} \\
 & & \\
\multirow{2}{*}{$A_{\mu}(f) \le 1.3\times 10^{-15}$} & \multirow{2}{*}{$A_{\mu}(f) \lesssim  2.1\times 10^{-15}$} & \multirow{2}{*}{$8.7\times 10^{-7}$~Hz $\le f \le 8.3\times 10^{-4}$~Hz}  \\
 & & \\
\hline\hline
\end{tabular}
\caption{Summary of constraints on the powers of fractional oscillations in $\alpha$ and $\mu$ produced in this paper, expressed as Fourier spectrum amplitude detection thresholds, $A_X(f)$, at the $1 \sigma$ and $5 \sigma$ significance level.}
\label{tab:oscillation constraints}
\end{table}

These constraints on sinusoidal oscillations can be further interpreted in the case of specific models such as scalar dark matter, which is the focus of Section \ref{Sec:DM}.

\section{Model-independent constraints}
\label{Sec:modind_limits}

As emphasized before, the main strength of our new theoretical description of a time evolution of constants is that it enables us to set model-independent constraints on the time variation of these constants. Independent of the shape of the function that describes the time variation, we now set some limits which can then trivially be interpreted in specific models.

From Eq.~\eqref{freqs}, a transition frequency
$\nu$ may be parametrized as~\cite{Flambaum:2004tm}
\begin{align} 
\label{nuu}
\nu = (\text{const.})\;(cR_\infty)\;\alpha^{K_{\alpha}}(m_e/\Lambda_{\rm QCD})^{K_\mu}
(m_{q}/\Lambda_{\rm QCD})^{K_q},
\end{align}
where $K_\alpha, K_\mu$, and $K_q$ are sensitivity coefficients
characteristic of the transition $\nu$ and $m_q \equiv (m_u+m_d)/2$. 
The quark coefficient $K_q = k_q^{M_N} + k_q^{g_N}$ 
parametrizes changes in the
nucleon mass $\delta M_N/M_N = k_q^{M_N}(\delta m_q/m_q)$
and the nuclear magnetic moment $\delta g_{N}/g_{N}
= k_q^{g_N}(\delta m_q/m_q)$ in terms of $m_q$ variations.
The ratio of two frequencies $r=\nu_1/\nu_2$
is independent of the dimensionful constants and varying Eq.~\eqref{nuu}
with respect to $\alpha, m_e/\Lambda_{\text{QCD}}$, and
$m_q/\Lambda_{\text{QCD}}$ noting~\eqref{variations} gives 
\begin{align}
\label{deltar}
\frac{\delta r}{r} = \left(\Delta K_{\alpha}d_\gamma^{(n)} + 
\Delta K_\mu(d_{m_e}^{(n)} - d_g^{(n)}) 
+ \Delta K_q(d_{q}^{(n)} - d_g^{(n)})\right)(\kappa\phi)^{n},
\end{align}
where $d^{(n)}_{q}$ is the mass-weighted mean-quark coupling. 
Using the published values of $K_\alpha^{\text{Yb}^+} = -5.95, 
K_\alpha^{\text{Sr}} = 0.06$, and $K_\alpha^{\text{Cs}} = 2.83$ from Ref.~\cite{Flambaum:2009VFCs} and 
$K_\mu^{\text{Cs}} = 1, K_q^{\text{Cs}} \approx 0.07$ from Ref.~\cite{Flambaum:2004tm} (and summarized in Tab.~\ref{tab:freq info}) results in the fractional frequency ratios
\begin{align}
&\left(\frac{\delta r}{r}\right)_{\text{Yb}^+/\text{Sr}} = -6.01d_\gamma^{(n)}(\kappa \phi)^n,\label{ratio1}\\
&\left(\frac{\delta r}{r}\right)_{\text{Sr}/\text{Cs}}
= -\left(2.77d_\gamma^{(n)} + d_{m_e}^{(n)}-d_g^{(n)} 
+ 0.07(d_{q}^{(n)} - d_g^{(n)})\right)(\kappa \phi)^n.\label{ratio2}
\end{align}
As can be observed in Fig.~\ref{fig:raw_MDEVs}, constraints from Yb$^+$/Sr measurements enable a bound on the coefficient $d_\gamma^{(n)}$ far below what is possible from Sr/Cs measurements. It is therefore appropriate to neglect $d_\gamma^{(n)}$ in Eq.~\eqref{ratio2}, which gives an effective coupling for the Sr/Cs measurements of
\begin{align}
\label{dSrCs}
d_{\text{Sr/Cs}}^{(n)} \approx d_{m_e}^{(n)}-d_g^{(n)} 
+ 0.07(d_{q}^{(n)} - d_g^{(n)}).
\end{align}

Comparing Eqs.~\eqref{ratio1} and~\eqref{ratio2} with the data in Fig.~\ref{fig:raw_MDEVs}, model-independent constraints can be placed on the magnitudes of the coupling strengths $d_\gamma^{(n)}$, $d_{\text{Sr/Cs}}^{(n)}$, and the instability of $\phi^n(t)$ over different timescales, $\sigma_{\phi^n}(\tau)$ given by
\begin{eqnarray}
\label{Eq_dgamma_model_indep}
 \kappa^n |d_\gamma^{(n)}| \sigma_{\phi^n}(\tau) \lesssim 2.3\times 10^{-16}/\sqrt{\tau /{\rm s}}\;,
\end{eqnarray}
 for timescales 60~s $\leq \tau \leq $ 30~000~s, and
 \begin{eqnarray}
 \label{Eq_dSrCs_model_indep}
 \kappa^n |d_{\text{Sr/Cs}}^{(n)}| \sigma_{\phi^n}(\tau) \lesssim 1.6\times 10^{-13}/\sqrt{\tau /{\rm s}}\;,
 \end{eqnarray}
for timescales 600~s $\leq \tau \leq $ 80~000~s, where $\sigma_{\phi^n}(\tau)$ is the Modified Allan Deviation of $\phi^{n}$, defined in Eq.~\eqref{mdevdef}, and the contribution from $(d_{q}^{(n)} - d_g^{(n)})$ is assumed to be subdominant. Note that these constraints do not assume any specific fundamental physics model, nor do they make any assumption about the functional form of $\phi^n(t)$. The constraints in the equations above are only valid for the specified values of $\tau$ explored in this work, and cannot constrain fluctuations on timescales outside this range.

The constraints on $\sigma_{\phi^n}(\tau)$ in Eqs.~\eqref{Eq_dgamma_model_indep} and~\eqref{Eq_dSrCs_model_indep} can be roughly interpreted as limits on the average magnitude of fluctuations in $\phi^n(t)$ between any two points in time $t$ and $(t+\tau)$.  For example, for two times separated by $\tau = 1~000$~s, the fluctuation $[\phi^n(t+\tau) - \phi^n(t)]$ should roughly satisfy 
$\kappa^n |d_\gamma^{(n)}| [\phi^n(t+\tau) - \phi^n(t)] < 7\times 10^{-18}.$

Only once the functional form of $\phi$ is specified is it possible for independent constraints to be placed on the couplings, which we explore in the special case of ultralight dark matter in the following section.

\section{Constraints on ultralight dark matter}
\label{Sec:DM}

One of the strongest cases for positing additional fundamental scalar fields is the problem of dark matter. Particle dark matter in the mass range $10^{-22}\;\text{eV} \lesssim m_{\phi} \lesssim 1\;\text{eV}$ is known as ultralight dark matter (ULDM), and in recent years significant efforts have been focused on detecting ULDM through apparent oscillations of fundamental constants. The upper bound $m_\phi \approx 1$\;eV occurs when the number density $n$ of bosons in the reduced de Broglie volume $(\lambda/2\pi)^3$ satisfies $n(\lambda/2\pi)^3 \gg 1$, resulting in a macroscopic phase-space occupation that exhibits Bose-Einstein condensation. Wavelengths $\lambda \sim \mathcal{O}(\text{kpc})$ span distances comparable to the smallest dwarf galaxy halos and imply a lower bound $m_\phi \approx 10^{-22}$\;eV~\cite{Gruzinov:2000fuzzyDM}. This bound also roughly coincides with the upper limit of dark matter being completely accounted for by $m_\phi$~\cite{Marsh:2015xka}, which is a common assumption in studies seeking to exclude couplings at a given confidence level.

The Standard Halo Model is assumed for the dark matter density and velocity profiles, where for coordinates centered on the galaxy $v_{\rm vir}\sim 10^{-3}$ is the virial speed (in natural units) of dark matter with isotropic distribution $\braket{\vec{v}_{\text{vir}}} = \vec{0}$. As a result of the solar system's motion through the dark matter halo at speeds comparable to $v_{\text{vir}}$, an Earth-based laboratory experiences a dark matter wind with $|\vec{k}| \approx m_\phi v_{\text{vir}} \ll m_\phi$ when neglecting subdominant corrections~\cite{MAYET20161}. In the dark-matter rest frame, oscillations are controlled by the rest mass $m_\phi = 2\pi f_\phi$ where $f_\phi$ is the Compton frequency in natural units. Oscillations are coherent in time for $\tau_c \sim 4\pi/(m_\phi v_{\rm vir}^2) \gtrsim 10^6 \ T_{c} \gg T_{\rm data}$ where the oscillation timescale $T_c = 1/f_\phi$ greatly exceeds the experimental timescale $T_{\rm data}$. As the coherence length $\lambda \sim 2\pi/(m_\phi v_{\rm vir})$ is larger than solar-system scales for all $f_\phi$ of interest, the scalar-field amplitude is approximately constant. Under these conditions, ULDM is described by a macroscopic, nearly constant-amplitude waveform oscillating at the underlying particle Compton frequency, up to small velocity corrections $\sim \mathcal{O}(v^2_{\text{vir}})$.

Measurements from the cosmic microwave background indicate dark matter was present in the early universe and is strongly constrained to be stable on  experimental timescales~\cite{PDG2022}. The standard theoretical treatment of $\phi$ as dark matter assumes cosmological evolution in a flat Friedmann–Lema\^{i}tre–Robertson–Walker universe where~\eqref{reducedEOM} follows with $\Gamma = 0$.
The field has an amplitude $\phi_0 = \sqrt{2\braket{\rho_{\phi}}}/m_\phi$, resulting from time-averaging the energy density $\rho_{\phi} = (\dot\phi^2 + m_\phi^2\phi^2)/2$ in the nonrelativistic limit. As all measurements presented in this paper were taken at a single location, the dynamics of $\phi$ are described by the solution 
\begin{equation}
\phi(t) \approx 
 \frac{\sqrt{2\braket{\rho_{\rm \phi}}}}{m_\phi}\cos(m_\phi t).
\end{equation}
Typically it is assumed that the scalar field comprises the entirety of the dark-matter density inferred at the solar galactocentric radius $R_0\simeq 8$\;kpc, or the ``local density" $\rho_{\rm DM}(R_0)\approx 0.3\;\text{GeV/cm}^3$~\cite{Read:2014qva}. This figure should be taken with caution and has a large influence on the constraints, as densities from solar-system objects including planetary ephemerides~\cite{Pitjev:2013sfa} and more recently asteroid data~\cite{Tsai:2022jnv} constrain  $\rho_{\text{DM}}\lesssim (10^4$-$10^6)\times \rho_{\text{DM}}(R_0)$. The immediate implications are on the sensitivity to the couplings in Eqs.~\eqref{ratio1} and \eqref{ratio2}, since for a generic coefficient (assuming the field saturates the total density) the sensitivity to $d_j^{(n)} \propto 1/\rho_{\text{DM}}^{n/2}$ and at best linear (quadratic) constraints would scale downward by $\sim 10^3$ ($\sim 10^6)$. However, they could also be substantially weakened. Constraints would need to be reconsidered in the case of multicomponent dark matter. For example, interactions between the different components could lead to nonzero decay  widths (i.e. $\Gamma\neq 0$), and the density contribution would be spread across the components $\rho_{\text{DM}} = \sum_i \braket{\rho_{\phi_i}}$. The extracted limits would be weaker and scaled by $\sqrt{\braket{\rho_{\phi_i}}/\rho_{\rm DM}}$.

\subsection{Scalar couplings}
Previous experiments monitoring electronic and microwave frequencies have resulted in constraints on the scalar couplings $d_\gamma^{(n)}, d_{m_f}^{(n)}$, and $d_g^{(n)}$ assuming dark matter~\cite{Arvanitaki15,Leefer:2015DM,Hees2016:DM,Kennedy:2020cavity,BACON:2021DM,Stadnik:2016DM-clocks,Stadnik:2015DM-VFCs,Stadnik15,Kalaydzhyan:2017jtv,Kobayashi:2022vsf}. In the present work, constraints may be extracted by using the amplitude spectra $A_X(f) = A_r(f)/|\Delta K|$ for frequency bin $f$ and relevant sensitivity coefficient $\Delta K$ from Fig.~\ref{fig:alphamu_spectra}. Identifying $f = f_\phi = m_\phi/2\pi$ for linear couplings and 
$f = 2f_\phi =m_\phi/\pi $ for quadratic couplings and noting Eq.~\eqref{deltar} gives
\begin{align} 
\label{excl}
d_j^{(n)} = \zeta_{f}^{(n)} A_X(f),
\end{align}
where $\zeta_{f}^{(1)} \approx 2.0\times 10^{16}f$/Hz, $\zeta_{f}^{(2)} \approx  (0.2\zeta_{f}^{(1)})^2$. Note that the effects of boosting from the dark matter rest frame to the laboratory frame, which introduces a broadening of the oscillation frequency $f_\phi\rightarrow f_\phi + m_\phi v_{\rm vir.}^2(4\pi)^{-1}$, have been included in $\zeta_{f}^{(1)}$. As a result, the sum over distinct field modes for measurement times $T\ll \tau_c$ reduces the sensitivity to the linear $n=1$ coefficients by a stochastic factor $\approx 3$~\cite{centers2020stochastic}.\footnote{To the best of our knowledge, the stochastic factor for quadratic couplings has not been calculated, so we set it equal to one.}
The data encompasses frequencies $10^{-6}\;\text{Hz} \lesssim f \lesssim 2\times 10^{-2}\;\text{Hz}$, corresponding to masses $4\times 10^{-21}\;\text{eV} \lesssim m_\phi \lesssim 8\times 10^{-17}\;\text{eV}$.

\begin{figure}[ht!]
\centering
\includegraphics[width=\textwidth]{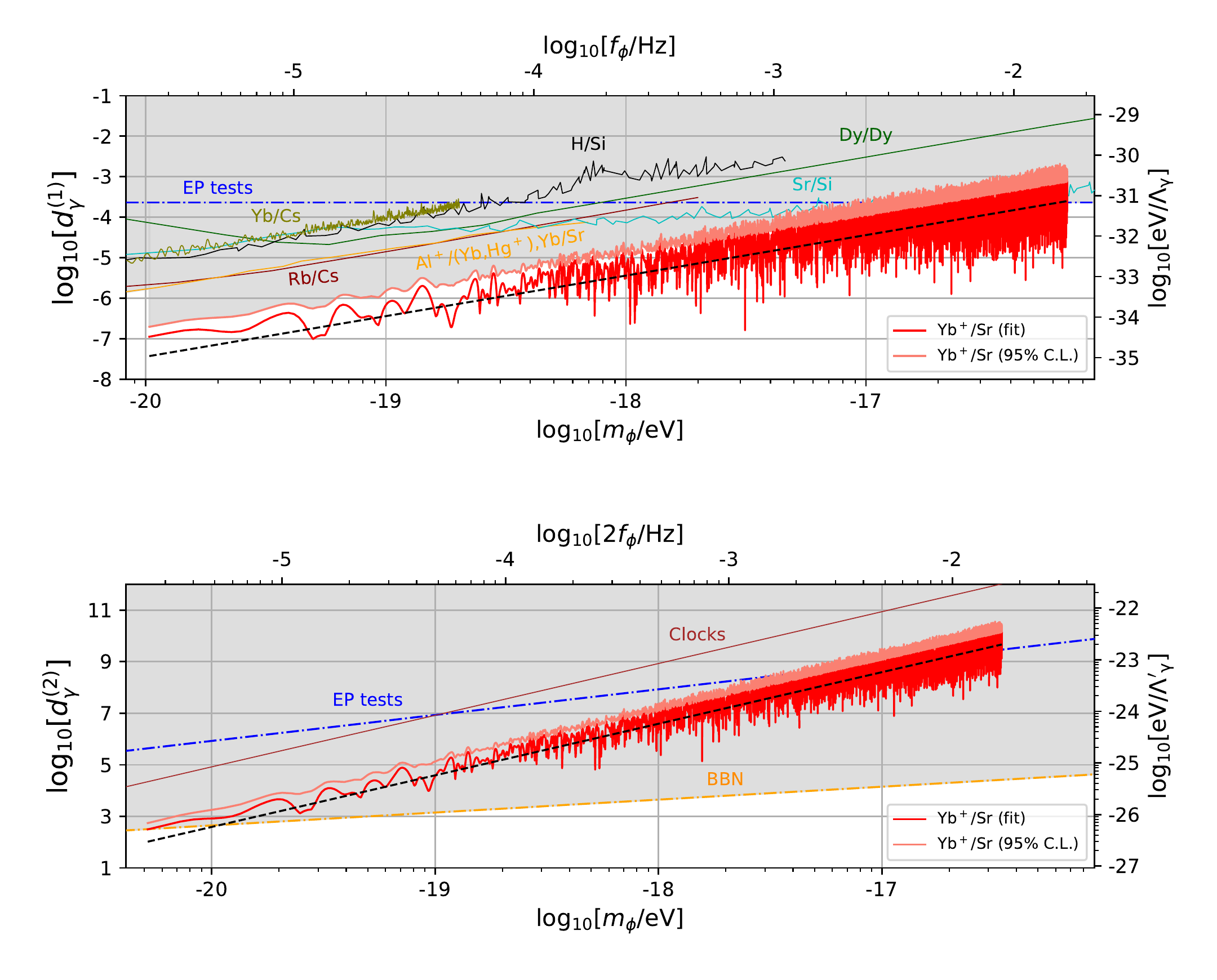}
\caption{Constraints on  $d_\gamma^{(1)}$ (top panel) and $d_\gamma^{(2)}$ (bottom panel). The best fit from Yb$^+$/Sr (red) along with expected noise level (black dashed line) and 95\% confidence level (C.L.) (light red) lines are displayed. Comparisons with constraints on linear couplings include Rb/Cs clocks~\cite{Hees2016:DM}, combined Al$^+$/(Yb, Hg$^+$), Yb/Sr clocks~\cite{BACON:2021DM}, Yb/Cs clocks~\cite{Kobayashi:2022vsf},  Sr/Si and H/Si cavity comparisons~\cite{Kennedy:2020cavity}, Dy/Dy spectroscopy~\cite{Leefer:2015DM} and EP tests~\cite{RotWash:1999,EotWash:2008,MICROSCOPE:2017A,MICROSCOPE:2017B}. Comparisons with constraints on quadratic couplings include clocks~\cite{Stadnik:2015DM-VFCs,Stadnik:2016DM-clocks}, EP tests~\cite{Hees:2018DM} and BBN~\cite{Stadnik:2015DM-VFCs}.}
\label{dgamma_lin_quad}  
\end{figure}

\begin{figure}[ht!]
\centering
\includegraphics[width=\textwidth]{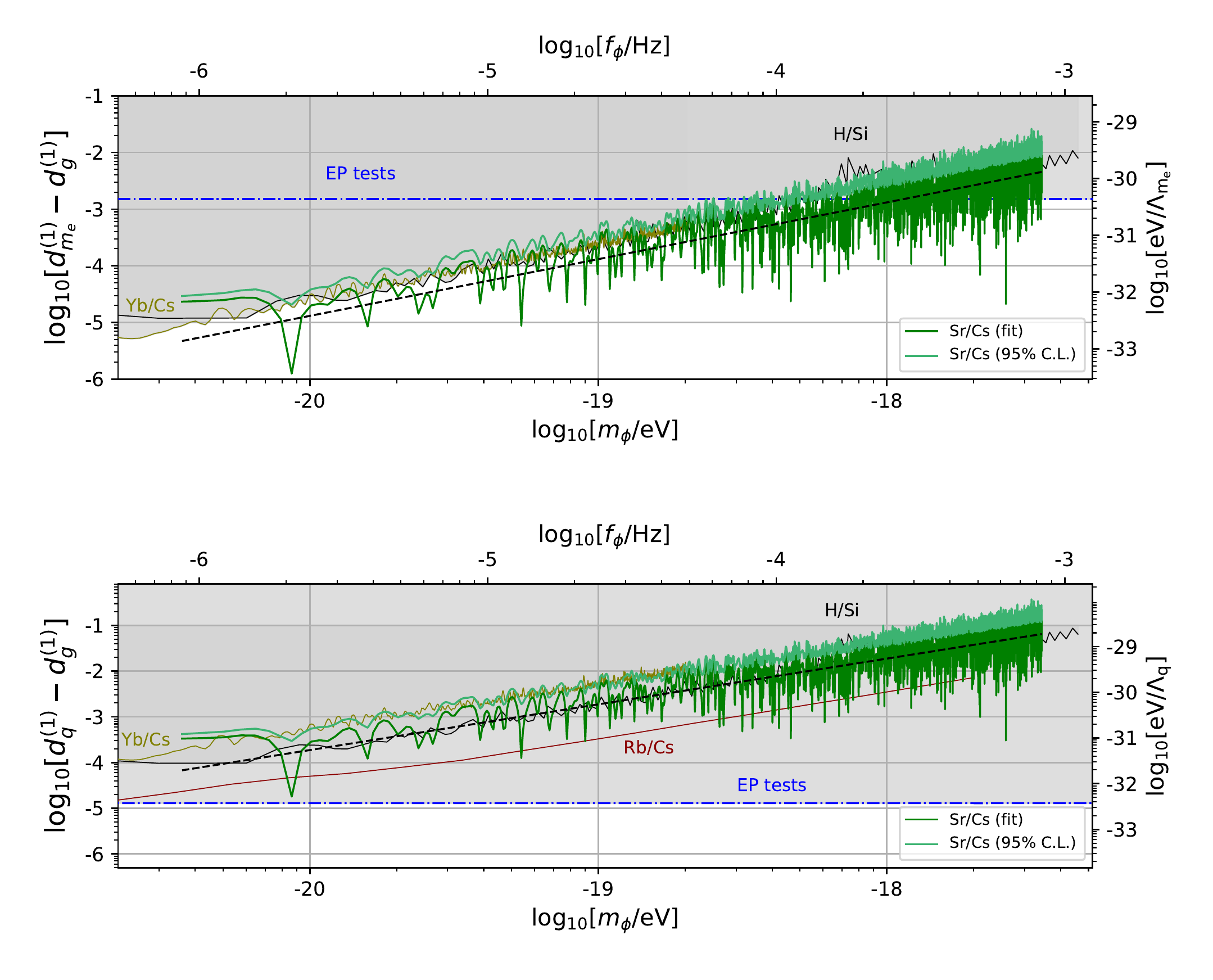}
\caption{Constraints on $d_{m_e}^{(1)}-d_{g}^{(1)}$ (top panel) and $d_{m_q}^{(1)}-d_{g}^{(1)}$ (bottom panel). The best fit from Sr/Cs (green) along with expected noise level (black dashed line) and 95\% C.L. (light green) lines are displayed, including comparisons with EP tests and Rb/Cs~\cite{Hees2016:DM}, Yb/Cs~\cite{Kobayashi:2022vsf}, and H/Si~\cite{Kennedy:2020cavity}.}
\label{dq_lin}
\end{figure}

\begin{figure}[ht!]
\centering
\includegraphics[width=\textwidth]{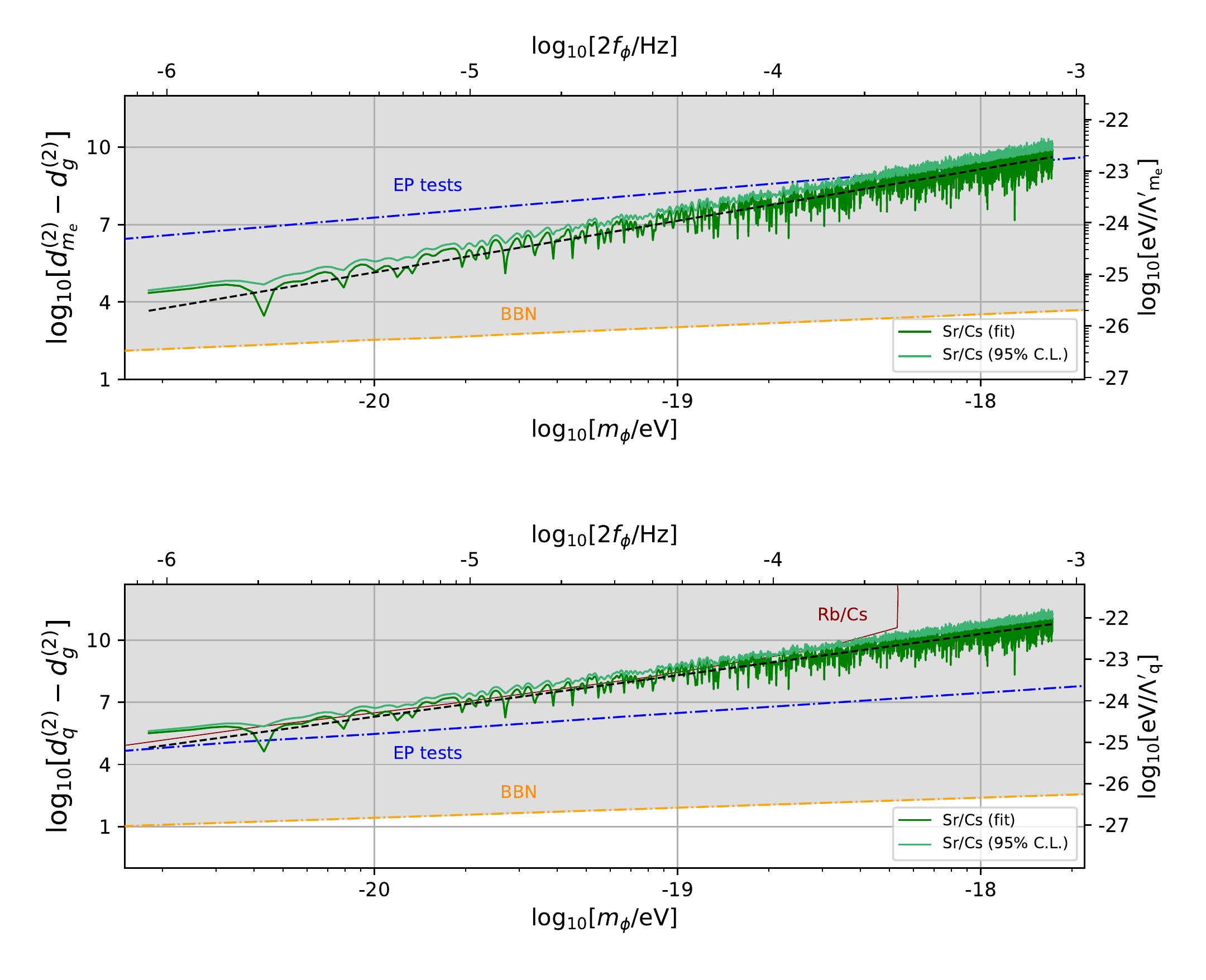}
\caption{Constraints on $d_{m_e}^{(2)}-d_{g}^{(2)}$ (top panel) and $d_{m_q}^{(2)}-d_{g}^{(2)}$ (bottom panel). The best fit from Sr/Cs (green) along with expected noise level (black dashed line) and 95\% C.L. (light green) lines are displayed, including comparisons with EP tests and Rb/Cs~\cite{Hees:2018DM}, and BBN~\cite{Stadnik:2015DM-VFCs,Stadnik:2016DM-clocks}.}
\label{dq_quad}
\end{figure}

\begin{figure}[ht!]
\centering
\includegraphics[width=0.9\textwidth]{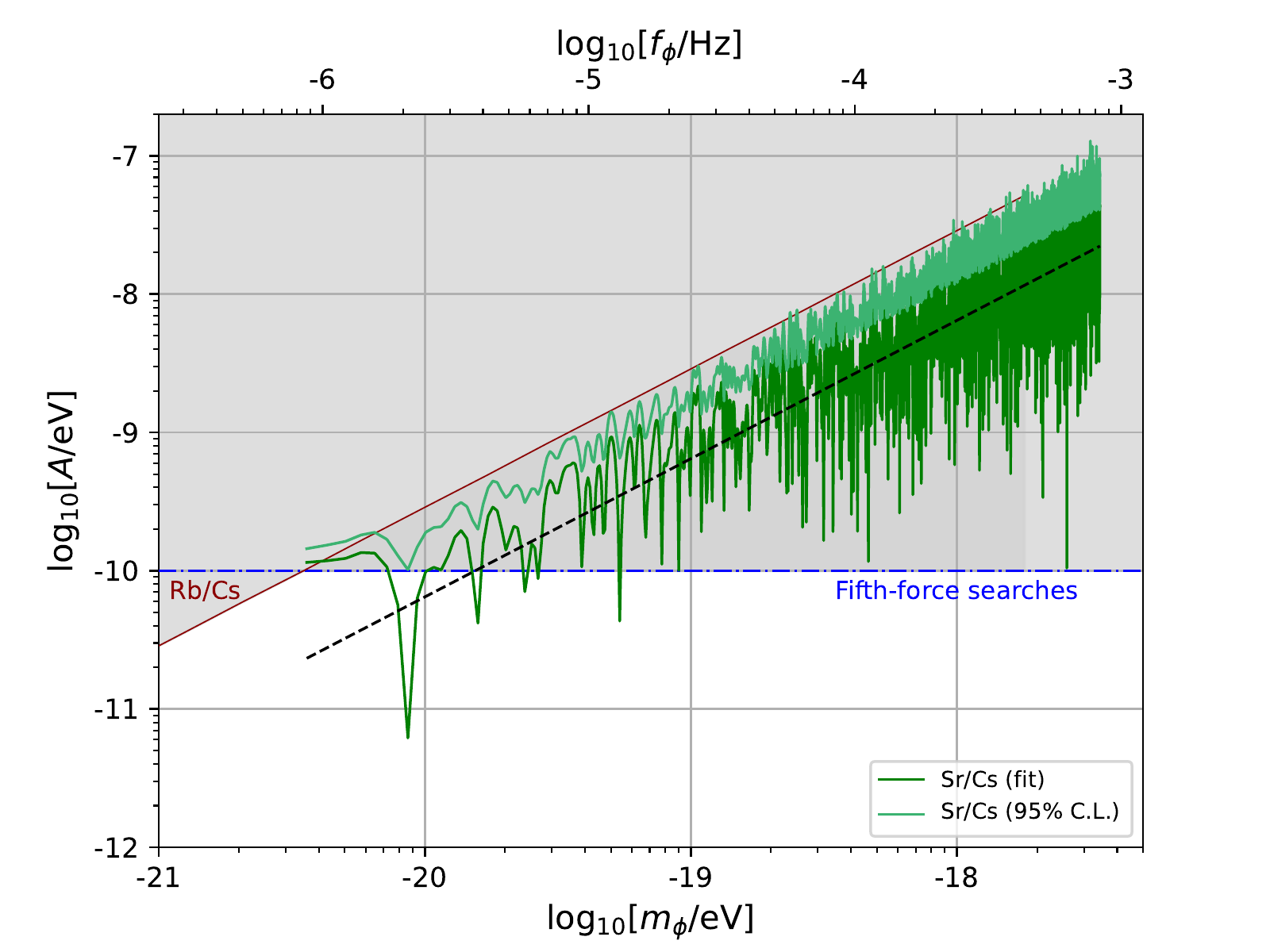}
\caption{Constraints on Higgs coupling parameter $A$. The best fit from Sr/Cs (green) along with expected noise level (black dashed line) and 95\% C.L. (light green) lines are displayed, including comparisons with Rb/Cs~\cite{Stadnik:2016DM-clocks}
and fifth-force searches~\cite{Pospelov2010}.}
\label{A_Higgs}
\end{figure}

The results for linear and quadratic scalar couplings discussed in Sec.~\ref{Sec:2} are presented for the case of ULDM in Figs.~\ref{dgamma_lin_quad}-\ref{A_Higgs}. Note that some works assume the slightly larger estimate $\rho_{\text{DM}}\approx 0.4\;\text{GeV/cm}^3$, which for purposes of comparison here amounts to a negligible difference.
For all figures, the right axes compare $\kappa^n d_j^{(n)} = 1/\Lambda_j^n$ for
a generic dimensionful scale $\Lambda_j$. We choose the scale to be identified with $\Lambda_j, \Lambda_j'$ for linear and quadratic couplings, respectively.
For the coupling $d_\gamma^{(1)}$ in Fig.~\ref{dgamma_lin_quad}, a new exclusion region up to roughly an order of magnitude improvement over previously published work for $10^{-20}\;\text{eV} \lesssim m_\phi \lesssim 10^{-17}\;\text{eV}$ is observed.  This improved sensitivity is mostly explained by the difference in $K$ factors between, e.g., Rb/Cs where $|\Delta K^{\text{Rb/Cs}}_\alpha| \approx 0.5$ is roughly an order of magnitude smaller than $|\Delta K^{\text{Yb$^+$/Sr}}_\alpha| \approx 6$. The extracted bounds also essentially surpass E\"ot-Wash~\cite{RotWash:1999,EotWash:2008} and MICROSCOPE~\cite{MICROSCOPE:2017A,MICROSCOPE:2017B} equivalence-principle (EP) tests assuming a light dilaton. We note that recent experimental results~\cite{PTB2023}, also using Yb$^+$ and Sr clocks, have claimed even tighter constraints than extracted here on the linear photon coupling $d_\gamma^{(1)}$ in this mass range.

For the quadratic coupling $d_\gamma^{(2)}$, a similar trend with respect to clock studies~\cite{Stadnik:2015DM-VFCs,Stadnik:2016DM-clocks} using Rb/Cs~\cite{Hees2016:DM} and $^{164}$Dy/$^{162}$Dy spectroscopy~\cite{Leefer:2015DM} persists and using EP-test results to extract bounds on quadratic couplings~\cite{Hees:2018DM} shows greater sensitivity $\gtrsim 4\times 10^{-18}$\;eV. We also include constraints from big bang nucleosynthesis (BBN)~\cite{Stadnik:2015DM-VFCs} which surpass the sensitivity of other experiments in the probed mass range.

To the best of our knowledge, the only previously published clock-based studies to account for stochastic degradation factor $\approx 3$ were those performed by the BACON collaboration with Al$^+$/(Yb, Hg$^+$) and Yb/Sr clock comparisons~\cite{BACON:2021DM}, JILA using clock-cavity Sr/Si and H/Si comparisons~\cite{Kennedy:2020cavity} and NMJI using Yb/Cs clock comparisons~\cite{Kobayashi:2022vsf}.
Note that EP tests do not rely on assumptions of the amplitude $\phi_0$ and thus the contribution of the scalar field to the dark matter abundance. Similarly, though BBN constraints use $\braket{\rho_\phi} = \rho_{\text{DM}}$, the field is non-oscillating with constant $\phi_0$ for $m_\phi \ll 10^{-16}$\;eV so coherence considerations are irrelevant. 

Limits on linear and quadratic $d_{m_e}^{(n)} - d_g^{(n)}$ and $d_{q}^{(n)} - d_g^{(n)}$ couplings are presented in Fig.~\ref{dq_lin} and Fig.~\ref{dq_quad}, respectively. The linear constraints are competitive and similar in shape and magnitude with H/Si, Rb/Cs and Yb/Cs comparisons over the range of data, however the sensitivity of Rb/Cs extends up to two orders of magnitude below EP tests around $m_\phi \approx 10^{-23}$\;eV. Note that Rb/Cs has no sensitivity to $d_{m_e}^{(n)} - d_g^{(n)}$. Regarding the quadratic constraints, Sr/Cs data probes a new region for clocks in the $d_{m_e}^{(2)} - d_g^{(2)}$ panel and displays roughly two orders of magnitude more sensitivity than EP tests at the low-mass range. Despite this, BBN still dwarfs the sensitivity by comparison and both EP tests and BBN encompass the range probed for $d_{q}^{(2)} - d_g^{(2)}$.

The final scalar constraints are extracted on the parameter $A$ from the scalar-Higgs interaction and are presented in Fig.~\ref{A_Higgs}. The simplest ($n=1$) linear couplings have garnered theoretical attention since they can emerge from the technically natural operator $\mathcal{L}_{\phi H} = -A\phi H^\dagger H$ for Higgs doublet $H$~\cite{Pospelov2010}. The sensitivity coefficient is
\begin{align}
K_H = \frac{\alpha}{2\pi}K_\alpha - (1-b)K_\mu - 1.05(1-b)K_q,
\end{align}
where $b\sim 0.2-0.5$ is a dimensionless factor in the Higgs-nucleon Yukawa coupling $g_{hNN} = bm_N/v$, where $v \approx 246$\;GeV is the Higgs vacuum expectation value and where the nucleon mass $m_N = (m_p + m_n)/2 \approx 0.94$\;GeV. To easily compare with existing Rb/Cs limits, we choose $b=0.2$ and use the relevant Sr/Cs values from Table 1 of Ref.~\cite{Stadnik:2016DM-clocks}. Using the Sr/Cs spectrum and noting $\kappa d_H \leftrightarrow A/m_h^2$ for Higgs mass $m_h \approx 125$\;GeV in~\eqref{excl} produces the limit. For the ratios considered here, the bulk of the sensitivity  comes from $(1-b)K_\mu$ as the electromagnetic portion is suppressed by an additional factor of $\alpha$ (from  radiative corrections) and the sensitivity of quark contributions to $m_N, g_N$ are suppressed relative to $m_e/m_N$ by around an order of magnitude. This implies, e.g., optical-optical Yb$^+$/Sr and microwave-microwave Rb/Cs ratios have weaker sensitivity to $A$ relative to optical-microwave ratios. It is worth noting that $b=0.5$ gives almost no sensitivity to Rb/Cs. Accordingly, Sr/Cs comparisons have more robust potential for constraining $A$, however, as the existing Rb/Cs data set is longer, competitive limits with respect to fifth-force searches exist in the range $10^{-24} \lesssim m_\phi \lesssim 10^{-20}$. Note again that the current Rb/Cs limits do not include the stochastic degradation factor $\approx 3$, which would shift the displayed curve upwards accordingly.

\subsection{Pseudoscalar couplings}
\begin{figure}
\centering
\includegraphics[width=0.9\textwidth]{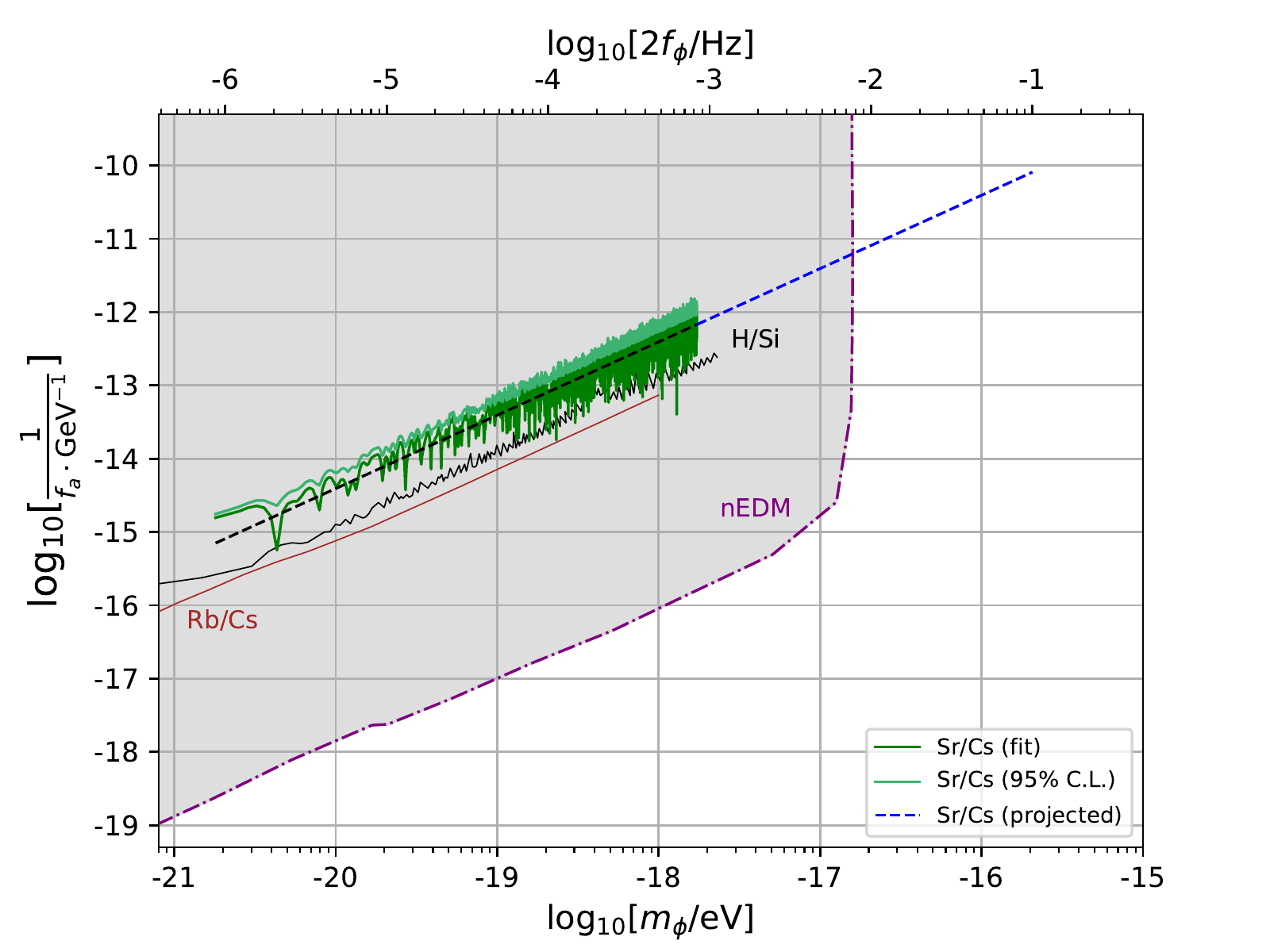}
\caption{Constraints on the QCD axion-like coupling $f_a$. The best fit from Sr/Cs (green) along with expected (black dashed line) and projected (blue dashed line) noise levels, and 95\% C.L. (light green) lines are displayed. 
Comparisons include Rb/Cs~\cite{Hees2016:DM}, H/Si~\cite{Kennedy:2020cavity} and nEDM~\cite{Abel:2017rtm}.}
\label{axionlimits}
\end{figure}

Recently Kim and Perez~\cite{Kim:2022ype} highlighted that atomic clocks 
can also provide complementary and competitive constraints 
on an axion-like field $a$ coupled to gluons:
\begin{equation}
\label{axL}
\mathcal{L}_a = \frac{g_s^2}{32\pi^2}\frac{a}{f_a} G_{\mu\nu}^b\widetilde{G}^{b\mu\nu},
\end{equation}
where $g_s$ and $f_a$ are the strong coupling and axion decay constant, respectively. 
As a result, the square of the pion mass undergoes small oscillations quadratic in the field
\begin{equation}
\label{pishift}
\frac{\delta m_\pi^2 }{m_\pi^2} = - \frac{m_u m_d}{2(m_u + m_d)^2} \theta^2,
\end{equation}
where $m_u, m_d$ are the up and down quarks and $\theta = a/f_a$. The nucleon mass $m_N(\theta)$ and the nucleon $g$-factor $g_N(\theta)$ (and hence \textit{nuclear} $g$-factor $g(\theta)$) have an inherent dependence on $m_\pi(\theta)$ as parametrized by chiral perturbation theory~\cite{Gasser:1987rb,SchererChiPT,ChiPT1,FlambaumTedesco}. Similar to quadratic scalar couplings in the ultralight range, the oscillatory component of $\theta$ is given by
\begin{align}
\label{theta2}
\theta^2 (t) = \frac{\rho_{\text{DM}}}{m_a^2 f_a^2}\cos(2m_a t),
\end{align}
where here $\rho_{\rm DM} = 0.4$\;GeV/cm$^3$, $m_a \not\propto f_a^{-1}$ in general and assuming $a$ saturates the local density. 
The variation of a microwave transition frequency~\eqref{freqs} may be expressed as variations in $g(\theta)$ and $m_p$ using Eqs.~\eqref{pishift} and~\eqref{theta2}, giving~\cite{PDG2022}
\begin{align}
\frac{\delta \nu_{\rm MW}}{\nu_{\rm MW}} \approx  -0.11\left[\frac{\partial \ln g(\theta)}{\partial \ln m_\pi^2}
- \frac{\partial \ln m_p}{\partial \ln m_\pi^2}
\right]\frac{\rho_{\text{DM}}}{m_a^2 f_a^2}\cos(2m_a t).
\label{deltaMW}
\end{align}
Comparing with another microwave or optical standard and identifying $m_a = \pi f$ for signal frequency $f$ from an amplitude spectrum yields the relation
\begin{align}
\label{finv}
\frac{1}{f_a\cdot\text{GeV}^{-1}} = 10^{-10}\sqrt{\frac{m_{15}^2}{c_r\cdot 10^{-15}}\bigg|\frac{\delta r}{r}\bigg|},
\end{align}
where $m_{15} \equiv m_a/(10^{-15}\;\text{eV})$ and $c_r$ is a constant. In~\cite{Kim:2022ype} microwave-microwave (Rb/Cs, $c_r\approx 10^{-1}$) and microwave-optical (H/Si, $c_r \approx 1$) comparisons are studied. One may also consider Sr/Cs, where due to the weak dependence of Sr on nuclear quantities $|\delta r/ r| \approx \delta \nu_\text{Cs}/\nu_\text{Cs}$ and $c_r\approx 2\times 10^{-2}$, including effects of stochasticity. From Eq.~\eqref{finv}, we plot the Sr/Cs limits along with Rb/Cs and H/Si from~\cite{Kim:2022ype} as well as include 
the constraints from the neutron electric dipole moment (nEDM) in Fig.~\ref{axionlimits}. 

Perhaps most compelling are the
high-frequency range constraints achievable on the axion-like coupling in Fig.~\ref{axionlimits} (see blue dashed line). In order to measure and remove density-dependent effects, the cesium fountain interleaves regular measurements with samples recorded with much higher atom densities. For the data presented here, this leads to a Cs clock cycle time of 600~s, corresponding to a maximum (Nyquist) frequency of $833$ $\mu$Hz. However, the cesium fountain could run with lower cycle times, and if the cesium fountain were operated without compensating density-dependent effects, cycle times of $\approx 5$~s could be achieved, allowing Fourier frequencies up to $\approx 0.1$~Hz to be probed, though at the expense of stability at longer times. This would yield constraints over an additional order of magnitude in frequency space outside of the large excluded nEDM region.
This contrasts with the projections from future experimental proposals, demonstrating that
existing atomic-clock capabilities can provide competitive constraints in axion physics. Taken
together, subsequent studies based on future data campaigns are clearly motivated.

\section{Discussion and Conclusion}
\label{Sec:conclusions}
In this work, we have presented a theoretical framework to describe the time variation of fundamental constants in a model-independent way. This approach demonstrates that a realistic model for the time variation of fundamental constants has many free parameters.

Using data acquired from atomic clocks operating at optical ($^{87}$Sr, $^{171}$Yb$^{+}$) and microwave ($^{133}$Cs) frequencies, we constrain the instability of fractional changes in $\alpha$ to be $\sigma_{(\Delta\alpha/\alpha)}(\tau)~\le 2.3\times 10^{-16}/\sqrt{\tau/{\rm s}}$ for averaging times 60~s~$ < \tau <$ 30~000~s, and we constrain the instability of fractional changes in $\mu$ to be $\sigma_{(\Delta\mu/\mu)}(\tau) \le 1.6\times 10^{-13}/\sqrt{\tau/{\rm s}}$ for averaging times $600$~s~$< \tau < 80~000$~s.  The theoretical framework then allows us to place constraints on combinations of a new scalar field $\phi^n(t)$ and the different interaction strengths $d_j^{(n)}$ coupling that field to other particles: photons, electrons, quarks and gluons. These constraints are independent of the underlying physics and the functional form of $\phi^n(t)$.

As an example of a specific model, we studied ultralight dark matter couplings to matter and presented new constraints on low-dimension dilaton-like operators. The limits on $d_\gamma^{(1)}$ from Yb$^+$/Sr data exclude a new region of parameter space for masses $10^{-20}{\rm eV} \lesssim m_\phi \lesssim 10^{-17}{\rm eV}$, as shown in Fig.~\ref{dgamma_lin_quad}. We also refer the reader to new experimental results~\cite{PTB2023} using Yb$^+$ and Sr clocks, which claim even tighter constraints on $d_\gamma^{(1)}$.

In the future, the limits on most of the parameters presented in this paper could readily be extended into both higher and lower frequency regions.  Higher frequency regions could be accessed by operating the clocks with shorter measurement cycles, and lower frequency regions could be accessed by recording longer data sets.  Furthermore, there are clocks being developed that are more sensitive to variations of fundamental constants, such as certain highly-charged ion species~\cite{Kozlov2018} or molecular clocks~\cite{Hanneke2021}.  Frequency ratios between these new types of optical clock could place significantly lower bounds on many of the coupling strengths between new scalar fields and particles of the Standard Model~\cite{Barontini2022}.

{\it Acknowledgments}: This work was carried out as part of the QSNET project within the Quantum Technologies for Fundamental Physics Programme, with support from the Science and Technology Facilities Council (grant numbers ST/T00102X/1, ST/T00598X/1 and ST/T006048/1).  We also acknowledge the support of the UK government department for Business, Energy and Industrial Strategy through the UK National Quantum Technologies Programme. We thank Abhishek Banerjee and Tejas Deshpande for providing exclusion-region results from previous experiments in Figs.~\ref{dgamma_lin_quad} and~\ref{dq_lin}, respectively, and Takumi Kobayashi for providing Yb/Cs data. X.C. would like to dedicate this work to the memory of Professor Harald Fritzsch who passed away on the $16^{th}$ of August 2022. 

{\it Data Availability}: The data that support the findings of this study are available upon request from the authors.

\appendix
\section{Appendix A} \label{Ap1}

The scalar field $\phi$ can couple to quarks or gluons linearly ($n=1$)
\begin{equation}
	{\cal L}=\kappa  \phi \left (\frac{d^{(1)}_g}{4} G_{\mu\nu}G^{\mu\nu} -d^{(1)}_{m_f} m_f \bar \psi_f \psi_f \right ),
\end{equation}
with $\kappa=\sqrt{4 \pi G}$, but we could also have quadratic couplings ($n=2$)
\begin{equation}
	{\cal L}=\kappa^2  \phi^2 \left (\frac{d^{(2)}_g}{4} G_{\mu\nu}G^{\mu\nu} -d^{(2)}_{m_f} m_f \bar \psi_f \psi_f \right).
\end{equation}
One can  disentangle  the different parameters $\phi_0$ and $d_j$. By considering experiments sensitive to $\alpha$, $\mu=m_e/m_p$ or $\alpha_s$, one could in principle measure $\phi_0$ and some of the $d_j$ independently. Furthermore, in general there could be several scalar fields. Some could couple to photons others to gluons. To be very clear, couplings may not be universal.

The mass of the proton $m_p$ is mainly sensitive to the time dependence in the QCD coupling constant. Remember that the QCD scale is given by $\Lambda_{\rm QCD}=\mu_r \exp(2 \pi/(\alpha_s(\mu_r)))^{1/b_3}$ with $b_3=-7$ in the Standard Model and where $\mu_r$ is the energy scale at which $\alpha_s$ is measured.
Neglecting a possible change in the quark masses,  the proton mass $m_p$ is proportional to $\Lambda_{\rm QCD}$. Using the renormalization group equation for $\alpha_s$ we find
\begin{equation}
\frac{\dot m_p}{m_p} \approx\frac{\dot \Lambda_{\rm QCD}}{\Lambda_{\rm QCD}}=-\frac{2\pi}{\beta} \frac{\dot \alpha_s}{\alpha_s^2}, \end{equation} 
where in the linear case, we have in the underdamped regime
\begin{eqnarray}
\frac{\dot \alpha_s}{\alpha_s}=-\frac{\kappa d^{(1)}_g \phi_0}{2} \exp{\left(-\frac{ \Gamma t}{2}\right)} \bigg(  \Gamma \cos\left( \theta - t \omega_{d}\right) + 2 \omega_{d} \sin\left( \theta - t \omega_{d} \right) \bigg)
\end{eqnarray}
and similarly again in the underdamped regime
\begin{eqnarray}
\frac{\dot \alpha_s}{\alpha_s}=-\frac{\kappa^2 d^{(2)}_g \phi_0^2}{2} \exp{\left(-\frac{ \Gamma t}{2}\right)} \bigg(  \Gamma \cos\left( \theta - t \omega_{d}\right) + 2 \omega_{d} \sin\left( \theta - t \omega_{d}\right) \bigg)
\end{eqnarray}
for the quadratic case.


\begin{thebibliography}{99}

\bibitem{Dirac:1937ti} 
P.~A.~M.~Dirac, The Cosmological Constants, Nature {\bf 139}, 323 (1937).

\bibitem{Dirac:1938mt}
P.~A.~M.~Dirac, A new basis for cosmology, Proc. Roy. Soc. Lond. A {\bf 165}, 199 (1938).

\bibitem{Marciano:1983wy}
W.~J.~Marciano, Time Variation of the Fundamental ``Constants" and Kaluza-Klein Theories, Phys.~Rev.~Lett.~{\bf 52}, 489 (1984).

\bibitem{Kolb:1985sj}
E.~W.~Kolb, M.~J.~Perry, and T.~P.~Walker, Time variation of fundamental constants, primordial nucleosynthesis, and the size of extra dimensions, Phys.~Rev.~D~{\bf 33}, 869~(1986).

\bibitem{Barrow:1987sr}
J.~D.~Barrow, Observational limits on the time evolution of extra spatial dimensions, Phys.~Rev.~D~{\bf 35} 1805 (1987).

\bibitem{Maeda:1987ku}
K.~I.~Maeda, On time variation of fundamental constants in superstring theories, Mod.~Phys.~Lett.~A~{\bf 3}, 243 (1988).

\bibitem{Barr:1988xw}
S.~M.~Barr and P.~K.~Mohapatra, Changing coupling ``constants" and violation of the equivalence principle, Phys.~Rev.~D~{\bf 38}, 3011 (1988).

\bibitem{Uzan:2011}
J.~-P.~Uzan, Varying Constants, Gravitation and Cosmology,  Liv.~Rev.~Rel. {\bf 14}, 2 (2011).

\bibitem{Martins:2017yxk}
C.~J.~A.~P.~Martins, The status of varying constants: a review of the physics, searches and implications, Rep.~Prog.~Phys.~{\bf 80}, 126902 (2017).

\bibitem{Bekenstein}
J.~D.~Bekenstein, Fine-structure constant: Is it really a constant?, Phys.~Rev.~D~{\bf 25}, 1527 (1982).

\bibitem{Barrow:1998df}
J.~D.~Barrow and J.~Magueijo, Varying alpha theories and solutions to the cosmological problems, Phys.~Lett.~B~{\bf 443}, 104 (1998).

\bibitem{Barrow:1999is}
J.~D.~Barrow, Cosmologies with varying light speed, 
Phys.~Rev.~D~{\bf 59}, 043515 (1999).

\bibitem{Derevianko14}
A.~Derevianko and M.~Pospelov, Hunting for topological dark matter with atomic clocks, Nature Phys.~{\bf 10}, 933 (2014).

\bibitem{Stadnik14}
Y.~V.~Stadnik and V.~V.~Flambaum, Searching for Topological Defect Dark Matter via Nongravitational Signatures, Phys.~Rev.~Lett.~{\bf 113}, 151301 (2014).

\bibitem{Arvanitaki15}
A.~Arvanitaki, J.~Huang, and K.~V.~Tilberg, Searching for dilaton dark matter with atomic clocks, Phys.~Rev.~D~{\bf 91}, 015015 (2015).

\bibitem{NewHorizons}
D.~Antypas~{\it et al.}, New Horizons:~Scalar and Vector Ultralight Dark Matter, arXiv:2203.14915. 

\bibitem{Rosenband2008}
T.~Rosenband~{\it et al.}, Frequency ratio of Al$^+$ and Hg$^+$ single-ion optical clocks; metrology at the 17th decimal place, Science~{\bf 319}, 1808, (2008).

\bibitem{Godun2014}
R.~M.~Godun~{\it et al.}, Frequency ratio of two optical clock transitions in $^{171}{\mathrm{Yb}}^{+}$ and constraints on the time variation of fundamental constants, Phys.~Rev.~Lett.~{\bf 113}, 210801, (2014).

\bibitem{Huntemann2014}
N.~Huntemann~{\it et al.}, Improved limit on a temporal variation of ${m}_{p}/{m}_{e}$ from comparisons of ${\mathrm{Yb}}^{+}$ and Cs atomic clocks, Phys.~Rev.~Lett.~{\bf 113}, 210802, (2014).

\bibitem{Lange21}
R.~Lange~{\it et al.}, Improved limits for violations of local position invariance from atomic clock comparisons, Phys.~Rev.~Lett.~{\bf 126}, 011102 (2021).

\bibitem{Hees2016:DM}
A.~Hees~{\rm et al.}, Searching for an Oscillating Massive Scalar Field as a Dark Matter Candidate Using Atomic Hyperfine Frequency Comparisons, Phys.~Rev.~Lett.~{\bf 117}, 061301 (2016).

\bibitem{Kennedy:2020cavity}
C.~J.~Kennedy~{\it et al.}, Precision Metrology Meets Cosmology: Improved Constraints on Ultralight Dark Matter from Atom-Cavity Frequency Comparisons, Phys.~Rev.~Lett.~{\bf 125}, 201302 (2020).

\bibitem{BACON:2021DM}
BACON Collaboration, Frequency ratio measurements at 18-digit accuracy using an optical clock network, Nature~{\bf 591}, 564 (2021).

\bibitem{Kobayashi:2022vsf}
T.~Kobayashi~{\it et al.}, Search for Ultralight Dark Matter from Long-Term Frequency Comparisons of Optical and Microwave Atomic Clocks, Phys.~Rev.~Lett.~{\bf 129}, 241301 (2022).

\bibitem{PTB2023}
M.~Filzinger~{\it et al.}, Improved limits on the coupling of ultralight bosonic dark matter to photons from optical atomic clock comparisons, 
Phys.~Rev.~Lett.~{\bf 130}, 253001 (2023).

\bibitem{Wcislo2018}
P.~Wcisło~{\it et al.}, New bounds on dark matter coupling from a global
network of optical atomic clocks, Sci.~Adv.~{\bf 4}, eaau4869 (2018).

\bibitem{Roberts2020}
B.~M.~Roberts~{\it et al.}, Search for transient variations of the fine structure constant and dark matter using fiber-linked optical atomic clocks, New J.~Phys.~{\bf 22}, 093010 (2020).

\bibitem{Burgess:2020tbq}
C.~P.~Burgess, Introduction to Effective Field Theory, Cambridge University Press, 2020,
ISBN 978-1-139-04804-0, 978-0-521-19547-8
doi:10.1017/9781139048040.

\bibitem{Gasser:1982ap}
J.~Gasser and H.~Leutwyler, Quark masses, Phys. Rept. {\bf 87}, 77 (1982).

\bibitem{Carroll98}
S.~Carroll, Quintessence and the rest of the world, Phys.~Rev.~Lett.~{\bf 81}, 3067 (1998).

\bibitem{Martin08}
J.~Martin, Quintessence: a mini review,
Mod.~Phys.~Lett.~A~{\bf 23}, 1252 (2008).

\bibitem{Dvali:2001dd}
G.~R.~Dvali and M.~Zaldarriaga, Changing 
$\alpha$
 with Time: Implications for Fifth-Force-Type Experiments and Quintessence, Phys.~Rev.~Lett.~{\bf 87}, 091303 (2002).

\bibitem{Lee:2020zjt}
J.~G.~Lee~{\it et al.}, 
New Test of the Gravitational $1/r^2$ Law at Separations down to 52 $\mu$m, Phys.~Rev.~Lett.~{\bf 124}, 101101 (2020).

\bibitem{Calmet:2019frv}
X.~Calmet, Hidden sector and gravity, Phys.~Lett.~B~{\bf 801}, 135152 (2020).

\bibitem{Campbell:1994bf}
B.~A.~Campbell and K.~A.~Olive, 
Nucleosynthesis and the time dependence of fundamental couplings, Phys.~Lett.~B~{\bf 345}, 429 (1995).

\bibitem{polchinski1}
J.~Polchinski, String theory. Vol. 1: An introduction to the bosonic string, Cambridge University Press (2007).

\bibitem{polchinski2}
J.~Polchinski, String theory. Vol. 2: Superstring theory and beyond, Cambridge University Press (2007).

\bibitem{Calmet:2017czo}
X.~Calmet, Cosmological evolution of the Higgs boson’s vacuum expectation value, Eur.~Phys.~J.~C~{\bf 77}, 729 (2017).

\bibitem{Calmet:2019nfj}
X.~Calmet, Cosmological time evolution of the Higgs mass and gravitational waves, Int.~J.~Mod.~Phys.~A~{\bf 35}, 2040035 (2020).

\bibitem{Hill:1983xh}
C.~T.~Hill, Are there significant gravitational corrections to the unification scale?, Phys.~Lett.~B~{\bf 135}, 47 (1984).

\bibitem{Vayonakis:1993nn}
A.~Vayonakis, Planck scale corrections to gauge coupling unification, Phys.~Lett.~B~{\bf 307}, 318 (1993).

\bibitem{Calmet:2009uz}
X.~Calmet and S.~W.~Majee, Effective theory for dark matter and a new force in the dark matter sector, Phys.~Lett.~B~{\bf 679}, 267 (2009).

\bibitem{Calmet:2019jyz}
X.~Calmet, On searches for gravitational dark matter with quantum sensors, Eur.~Phys.~J.~Plus~{\bf 134}, 503 (2019).

\bibitem{Calmet:2020pub}
X.~Calmet and F.~Kuipers, Theoretical bounds on dark matter masses, Phys.~Lett.~B~{\bf 814}, 136068 (2021).

\bibitem{Calmet:2022bin}
X.~Calmet and N.~Sherrill, Implications of Quantum Gravity for Dark Matter Searches
with Atom Interferometers, Universe~{\bf 8}, 103 (2022).

\bibitem{Calmet:2001nu}
X.~Calmet and H.~Harald, The Cosmological evolution of the nucleon mass and the electroweak coupling constants, Eur.~Phys.~J.~C~{\bf 24}, 639 (2002).

\bibitem{Calmet:2002ja}
X.~Calmet and H.~Fritzsch, Symmetry breaking and time variation of gauge couplings, Phys.~Lett.~B~{\bf 540}, 173 (2002).

\bibitem{Calmet:2002jz}
X.~Calmet and H.~Fritzsch, Grand unification and time variation of the gauge couplings, arXiv:0211421, published in the proceedings of  the 10th International Conference on Supersymmetry and Unification of Fundamental Interactions (SUSY02), Hamburg, Germany, 17-23 June 2002.

\bibitem{Calmet:2006sc}
X.~Calmet and H.~Fritzsch, A time variation of proton-electron mass ratio and grand unification, Europhys.~Lett.~{\bf 76}, 1064 (2006).

\bibitem{Flambaum:1999VFCs-A}
V.~A.~Dzuba, V.~V.~Flambaum, and J.~K.~Webb, Space-Time Variation of Physical Constants and Relativistic Corrections in Atoms, Phys.~Rev.~Lett.~{\bf 82}, 888 (1999).

\bibitem{Flambaum:1999VFCs-B}
V.~A.~Dzuba, V.~V.~Flambaum, and J.~K.~Webb, Calculations of the relativistic effects in many-electron atoms and space-time variation of fundamental constants, Phys.~Rev.~A~{\bf 59}, 230 (1999).

\bibitem{Flambaum:2009VFCs}
V.~V.~Flambaum and V.~A.~Dzuba, Search for variation of the fundamental constants in atomic, molecular, and nuclear spectra, Can.~J.~Phys.~{\bf 87}, 25 (2009).

\bibitem{radii1}
A.~Banerjee~{\it et al.}, Oscillating nuclear charge radii as sensors for ultralight dark matter, arXiv:2301.1078.

\bibitem{radii2}
V.~V.~Flambaum and I.~B.~Samsonov, Fluctuations of atomic energy levels due to axion and scalar fields, arXiv:2302.11167.

\bibitem{Hobson20}
R.~Hobson~{\it et al.}, A strontium optical lattice clock with 1 {\texttimes} 10$^{-17}$ uncertainty and measurement of its absolute frequency, Metrologia~{\bf 57}, 065026 (2020).

\bibitem{Baynham2018Ytterbium}
C.~F.~A.~Baynham~{\it et al.}, Absolute frequency measurement of the $^2$S$_{1/2}$ $\rightarrow$ $^2$F$_{7/2}$ optical clock transition in $^{171}$Yb$^+$ with an uncertainty of $4 \times 10^{-16}$ using a frequency link to international atomic time, J.~Mod.~Opt.~{\bf 65(5-6)}, 585-591 (2018).

\bibitem{Szymaniec2016}
K.~Szymaniec~{\it et al.}, NPL Cs fountain frequency standards and the quest for the ultimate accuracy, J.~Phys.:~Conf.~Ser.~{\bf 723}, 012003 (2016).

\bibitem{Dzuba2008}
V.~A.~Dzuba, V.~V.~Flambaum, Relativistic corrections to transition frequencies of Ag I, Dy I, Ho I, Yb II, Yb III, Au I, and Hg II and search for variation of the fine-structure constant, Phys.~Rev.~A~{\bf 77}, 012515 (2008).

\bibitem{berengut}
J.~C.~Berengut, V.~A.~Dzuba, and V.~V.~Flambaum, Enhanced Laboratory Sensitivity to Variation of the Fine-Structure Constant using Highly Charged Ions, Phys.~Rev.~Lett.~{\bf 105}, 120801 (2010).

\bibitem{Angstmann2004}
E.~J.~Angstmann, V.~A.~Dzuba, and V.~V.~Flambaum, Relativistic effects in two valence-electron atoms and ions and the search for variation of the fine-structure constant, Phys.~Rev.~A~{\bf 70}, 014102 (2004).

\bibitem{FlambaumTedesco}
V.~V.~Flambaum and A.~F.~Tedesco, Dependence of nuclear magnetic moments on quark masses and limits on temporal variation of fundamental constants from atomic clock experiments, Phys.~Rev.~C~{\bf 73}, 055501 (2006).

\bibitem{Schioppo22}
M.~Schioppo~{\it et al.}, Comparing ultrastable lasers at $7\times 10^{-17}$ fractional frequency instability through a 2220 km optical fibre network, Nature~Commun.~{\bf 13}, 212 (2022).

\bibitem{BIPMfreqs}
https://www.bipm.org/en/publications/mises-en-pratique/standard-frequencies

\bibitem{NISTNote1337}
D.~B.~Sullivan~{\it et al.}, NIST technical note 1337: Characterization of clocks and oscillators, (1990).

\bibitem{NISTbooklet}
W.~Riley and D.~Howe, Handbook of Frequency Stability Analysis, (2008).

\bibitem{Rubiola2005}
E.~Rubiola, On the measurement of frequency and of its sample variance with high-resolution counters, Review of Scientific Instruments~{\bf 76}, 054703 (2005).

\bibitem{Dawkins2007}
S.~T.~Dawkins, J.~J.~McFerran, and A.~N.~Luiten, Considerations on the measurement of the stability of oscillators with frequency counters, IEEE Transactions on Ultrasonics, Ferroelectrics, and Frequency Control~{\bf 54}, 918 (2007).

\bibitem{Benkler2015}
E.~Benkler, C.~Lisdat, and U.~Sterr, On the relation between uncertainties of weighted frequency averages and the various types of Allan deviations, Metrologia~{\bf 52}(4), 565 (2015).

\bibitem{Rubiola2023}
E.~Rubiola and F.~Vernotte,
The companion of Enrico’s chart for phase noise and two-sample variances, IEEE Transactions on Microwave Theory and Techniques, (2023).

\bibitem{RileyGaps}
W.J.~Riley, Gaps, Outliers, Dead Time, and Uneven Spacing in Frequency Stability Data, Hamilton Technical Services.

\bibitem{Stable32}
W.J.~Riley, User Manual Stable32 Frequency Stability Analysis, Hamilton Technical Services.

\bibitem{karshenboim2000}
S.~G.~Karshenboim, Some possibilities for laboratory searches for variations of fundamental constants, Can.~J.~Phys.~{\bf 78}, 639 (2000).

\bibitem{Bretthorst2006}
L.~G.~Bretthorst, Frequency Estimation and Generalized Lomb-Scargle Periodograms, Statistical Challenges in Astronomy, 309 (2006).

\bibitem{Leefer:2015DM}
K.~V.~Tilburg~{\it et al.}, Search for Ultralight Scalar Dark Matter with Atomic Spectroscopy, Phys.~Rev.~Lett.~{\bf 115}, 011802 (2015). 

\bibitem{lomb1976}
N.~Lomb, Least-squares frequency analysis of unequally spaced data, \\Astrophys.~Space~Sci.~{\bf 39}, 447 (1976).

\bibitem{scargle1982}
J.~D.~Scargle, Studies in astronomical time series analysis. II-Statistical aspects of spectral analysis of unevenly spaced data, The Astrophysical Journal~{\bf 263}, 835 (1982).

\bibitem{VanderPlas2018}
J.~T.~VanderPlas, Understanding the Lomb–Scargle Periodogram, The Astrophysical Journal Supplement Series~{\bf 236}, 16 (2018).

\bibitem{Munteanu2016}
C.~Munteanu~{\it et al.}, Effect of data gaps: Comparison of different spectral analysis methods, Annales Geophysicae~{\bf 34}, 437 (2016).

\bibitem{groth1975}
E.~J.~Groth, Probability distributions related to power spectra, The Astrophysical Journal Supplement Series~{\bf 29}, 285 (1975). 

\bibitem{ivezic2014}
Z.~Ivezi{\'c}~{\it et al.}, Statistics, Data Mining, and Machine Learning in Astronomy, Princeton University Press, (2014).

\bibitem{Baluev2008}
R.~V.~Baluev, Assessing the statistical significance of periodogram peaks, Monthly Notices of the Royal Astronomical Society~{\bf 385}, 1279 (2008).

\bibitem{Flambaum:2004tm}
V.~V.~Flambaum~{\it et al.}, Limits on the temporal variation of the fine structure constant, quark masses and strong interaction from quasar absorption spectra and atomic clock experiments, Phys.~Rev.~D~{\bf 69}, 115006 (2004).

\bibitem{Gruzinov:2000fuzzyDM}
W.~Hu, R.~Barkana, and A.~Gruzinov, Fuzzy Cold Dark Matter: The Wave Properties of Ultralight Particles, Phys.~Rev.~Lett.~{\bf 85}, 1158 (2000).

\bibitem{Marsh:2015xka}
D.~J.~E.~Marsh, Axion cosmology, Phys.~Rept.~{\bf 643}, 1 (2016).

\bibitem{MAYET20161}
F.~Mayet~{\it et al.}, A review of the discovery reach of directional Dark Matter detection, Physics Reports~{\bf 627}, 1 (2016).

\bibitem{PDG2022}
R.~L.~Workman~{\it et al.} (Particle Data Group), Review of Particle Physics, Progress of Theoretical and Experimental Physics~{\bf 2022}, (2022).

\bibitem{Read:2014qva}
J.~I.~Read, The local dark matter density, J.~Phys.~G: Nucl.~Part.~Phys.~{\bf 41}, 063101 (2014).

\bibitem{Pitjev:2013sfa}
N.~P.~Pitjev and E.~V.~Pitjeva, Constraints on dark matter in the solar system, Astron. Lett.~{\bf 39}, 141 (2013). 

\bibitem{Tsai:2022jnv}
Y.~-D.~Tsai~{\it et al.}, New Constraints on Dark Matter and Cosmic Neutrino Profiles through Gravity, arXiv:2210.03749.

\bibitem{Stadnik:2016DM-clocks}
Y.~V.~Stadnik and V.~V.~Flambaum, Improved limits on interactions of low-mass spin-0 dark matter from atomic clock spectroscopy, Phys.~Rev.~A~{\bf 94}, 022111 (2016). 

\bibitem{Stadnik:2015DM-VFCs}
Y.~V.~Stadnik and V.~V.~Flambaum, Can Dark Matter Induce Cosmological Evolution of the Fundamental Constants of Nature?, Phys.~Rev.~Lett.~{\bf 115}, 201301 (2015).

\bibitem{Stadnik15}
Y.~V.~Stadnik and V.~V.~Flambaum, Searching for Dark Matter and Variation of Fundamental Constants with Laser and Maser Interferometry, Phys.~Rev.~Lett.~{\bf 114}, 161301 (2015).

\bibitem{Kalaydzhyan:2017jtv}
T.~Kalaydzhyan and Y.~Nan, Extracting dark matter signatures from atomic clock stability measurements, Phys.~Rev.~D~{\bf 96}, 075007 (2017).

\bibitem{centers2020stochastic}
G.~P.~Centers~{\it et al.}, Stochastic fluctuations of bosonic dark matter, Nature~Commun.~{\bf 12}, 7321 (2021).

\bibitem{RotWash:1999}
G.~L.~Smith~{\it et al.}, 
Short-range tests of the equivalence principle, Phys.~Rev.~D~{\bf 61}, 022001 (1999).

\bibitem{EotWash:2008}
S.~Schlamminger~{\it et al.}, Test of the Equivalence Principle Using a Rotating Torsion Balance, Phys.~Rev.~Lett.~{\bf 100}, 041101 (2008).

\bibitem{MICROSCOPE:2017A}
P.~Touboul~{\it et al.}, MICROSCOPE Mission: First Results of a Space Test of the Equivalence Principle, Phys.~Rev.~Lett.~{\bf 119}, 231101 (2017).

\bibitem{MICROSCOPE:2017B}
J.~Berg\'e~{\it et al.}, MICROSCOPE Mission: First Constraints on the Violation of the Weak Equivalence Principle by a Light Scalar Dilaton, Phys.~Rev.~Lett.~{\bf 120}, 141101 (2018).

\bibitem{Hees:2018DM}
A.~Hees~{\it et al.}, Violation of the equivalence principle from light scalar dark matter, Phys.~Rev.~D~{\bf 98}, 064051 (2018).

\bibitem{Pospelov2010}
F.~Piazza and M.~Pospelov, Sub-eV scalar dark matter through the super-renormalizable Higgs portal, Phys.~Rev.~D~{\bf 82}, 043533 (2010).

\bibitem{Kim:2022ype}
H.~Kim and G.~Perez, Oscillations of atomic energy levels induced by QCD axion dark matter, arXiv:2205.12988.

\bibitem{Gasser:1987rb}
J.~Gasser, M.~E.~Sainio, and A.~Svarc, Nucleons with Chiral Loops, Nucl.~Phys.~B~{\bf 307}, 779 (1988).

\bibitem{SchererChiPT}
S.~Scherer and M.~S.~Schindler, A Primer for Chiral Perturbation Theory, Lect.~Notes~Phys.~{\bf 830}, (2012).

\bibitem{ChiPT1}
M.~Hoferichter~{\it et al.}, Matching Pion-Nucleon Roy-Steiner Equations to Chiral Perturbation Theory, Phys.~Rev.~Lett.~{\bf 115}, 192301 (2015).

\bibitem{Abel:2017rtm}
C.~Abel~{\it et al.},
Search for Axionlike Dark Matter through Nuclear Spin Precession in Electric and Magnetic Fields, Phys.~Rev.~X~{\bf 7}, 041034 (2017).

\bibitem{Kozlov2018}
M.~G.~Kozlov~{\it et al.},
Highly charged ions: Optical clocks and applications in fundamental physics, Rev.~Mod.~Phys.~{\bf 90}, 045005 (2018).

\bibitem{Hanneke2021}
D.~Hanneke~{\it et al.}, Optical clocks based on molecular vibrations as probes of variation of the proton-to-electron mass ratio, Quantum~Sci.~Technol.~{\bf 6}, 014005 (2021).

\bibitem{Barontini2022}
G.~Barontini~{\it et al.},
Measuring the stability of fundamental constants with a network of clocks, EPJ Quantum Technol.~{\bf 9}, 12 (2022).


\end{thebibliography}
\end{document}